\documentclass[11pt]{article}
\usepackage{amsfonts}
\usepackage{amssymb}
\usepackage{mathrsfs}
\usepackage{graphicx}
\usepackage{amsmath,amsthm,amssymb,amscd}
\usepackage[all]{xy}
\usepackage{subfig}
\usepackage{hyperref}
\usepackage{multirow}
\usepackage{color}
\usepackage{float}
\usepackage{mathtools,slashed}
\usepackage{bbold}
\usepackage{bbm}
\usepackage{slashed}

\usepackage{amsxtra,graphics,epsfig,bm,tikz,xfrac,lscape}
\usetikzlibrary{decorations.pathmorphing}
\usetikzlibrary{decorations.markings}
\usetikzlibrary{arrows, decorations.markings, calc, fadings, decorations.pathreplacing, patterns, decorations.pathmorphing, positioning}

\usetikzlibrary{positioning,shapes}
\usetikzlibrary{chains}
\usetikzlibrary{arrows,fit,decorations.pathreplacing}
\tikzstyle{every picture}+=[remember picture]
\tikzstyle{na} = [baseline=-.5ex]

\newcommand\fft[2]{\frac{#1}{#2}}
\newcommand\ft[2]{{\textstyle\frac{#1}{#2}}}
\newcommand\nn{\nonumber}
\newcommand\Tr{\operatorname{Tr}}
\newcommand\sgn{\operatorname{sgn}}
\renewcommand{\Re}{\operatorname{Re}}

\def\eqa{\begin{eqnarray}}
\def\eqae{\end{eqnarray}}
\def\eq{\begin{equation}}
\def\eqe{\end{equation}}
\def\be{\begin{equation}}
\def\ee{\end{equation}}
\def\bea{\begin{eqnarray}}
\def\eea{\end{eqnarray}}
\def\ba{\begin{array}}
\def\ea{\end{array}}
\def\bd{\begin{displaymath}}
\def\ed{\end{displaymath}}

\def\Tr{{\rm Tr}}

\def\>{\rangle}
\def\<{\langle}

\def\Td{\mathrm{Td}}
\def\Li{\operatorname{Li}}

\numberwithin{equation}{section}

\definecolor{darkblue}{rgb}{0,0,0.5}
\definecolor{darkred}{rgb}{0.5,0,0}
\definecolor{darkgreen}{rgb}{0,0.5,0}
\definecolor{orange}{rgb}{0.9,0.58,0}

\begin{document}

\begin{titlepage}
\hfill LCTP-18-20

\vskip 1 cm
\begin{center}
{\large \bf Subleading Microstate Counting in the Dual to Massive Type IIA }\\

\vskip .7cm


\end{center}

\vskip .7 cm

\vskip 1 cm
\begin{center}
{ \large James T. Liu${}^a$, Leopoldo A. Pando Zayas${}^{a,b}$ and Shan Zhou${}^{c}$}
\end{center}
\vskip .4cm \centerline{\it ${}^a$ Leinweber Center for Theoretical
Physics,   Randall Laboratory of Physics}
\centerline{ \it The University of
Michigan, Ann Arbor, MI 48109-1040}
\bigskip\bigskip

\centerline{\it ${}^b$ The Abdus Salam International Centre for Theoretical Physics}
\centerline{\it  Strada Costiera 11,  34014 Trieste, Italy}
\bigskip\bigskip

\centerline{\it ${}^c$ Department of Physics, University of California, Santa Barbara, CA 93106, USA }

\vskip 1 cm

\vskip 1.5 cm
\begin{abstract}
We study the topologically twisted index of a certain Chern-Simons matter theory with $SU(N)$ level $k$ gauge group on a genus $g$ Riemann surface times a circle. For this theory it is known that the logarithm of the  topologically twisted index grows as $N^{5/3}$ and that it matches the Bekenstein-Hawking entropy of certain magnetically charged asymptotically $AdS_4\times S^6$ black holes in massive type IIA supergravity. Through a combination of numerical and analytical techniques we study the subleading in $N$ structure. We demonstrate precise analytic cancellation of terms of orders  $N\log\,N$ and $N^{1/3}\log N$ and show numerical cancellation for terms of  order $N$.  As a result, the first subleading correction is of order $N^{2/3}$.  Furthermore, we provide evidence for the  presence of a term of the form $(g-1)(7/18) \log \,N$ which constitutes a microscopic prediction for the one-loop contribution coming from the massless gravitational degrees of freedom in the massive IIA black hole.

\end{abstract}

\vskip  1.5 cm

{\tt jimliu@umich.edu, lpandoz@umich.edu, shan\textunderscore zhou@ucsb.edu}

\end{titlepage}

\tableofcontents
\section{Introduction}

Understanding the microscopic origin of the Bekenstein-Hawking entropy  of black holes is one of the most challenging problems in theoretical physics; it constitutes a major test for any theory aspiring to wear the mantle of quantum gravity. Recently,  a microscopic explanation of the Bekenstein-Hawking entropy has been presented in the context of the AdS/CFT correspondence. Namely, the topologically twisted index of a Chern-Simons matter theory known as ABJM theory \cite{Aharony:2008ug} was shown to reproduce the leading term in the black hole entropy of certain magnetically charged asymptotically $AdS_4$ black holes \cite{Benini:2015eyy}.  Similar matches  have now been established for dyonic black holes \cite{Benini:2016rke}, black holes with hyperbolic horizons \cite{Cabo-Bizet:2017jsl} and asymptotically $AdS_4$ black holes in massive type IIA supergravity \cite{Benini:2017oxt,Azzurli:2017kxo,Hosseini:2017fjo}.  Moreover, some higher dimensional explorations have appeared recently \cite{Hosseini:2018uzp,Crichigno:2018adf}.

In this manuscript we focus on the topologically twisted index of $SU(N)$ at level $k$ Chern-Simons matter theory whose leading term, of order  $N^{5/3}$,  coincides with the entropy of magnetically charged, asymptotically $AdS_4\times S^6$ black holes  in massive type IIA theory \cite{Benini:2017oxt,Azzurli:2017kxo,Hosseini:2017fjo}.   The black holes in question were presented in \cite{Guarino:2017eag} as a payoff of the arduous work of obtaining $AdS_4$ gauged supergravity from the reduction of massive type IIA theory \cite{Guarino:2015jca,Guarino:2015qaa,Guarino:2015vca,Varela:2015uca}. 

Here, we mainly focus on the field theory side and develop an understanding of the sub-leading in $N$ corrections to the topologically twisted  index.  We  provide various pieces of evidence supporting the presence of a term of the form $(g-1)(7/18)\log N$ in the large $N$ expansion of the topologically twisted index. Although our work  in sub-leading corrections is strictly field theoretic, it is clearly motivated by the prospects of better understanding asymptotically $AdS_4$ black holes in ten-dimensional massive type IIA supergravity. In the context of the AdS/CFT correspondence one is to view the topologically twisted index as the microscopic description of such black holes. Therefore,  in this setup  the logarithmic in $N$ term should be considered as a prediction for the corresponding gravity result. More precisely, on the macroscopic (gravity) side these logarithmic corrections arise from one-loop contributions to the black hole partition function. Moreover, the logarithmic corrections arise only from loops of massless fields and from the range of loop momentum integration where the loop momenta remain much smaller than the Planck scale, thus creating an IR window into the microstates.

For supersymmetric extremal black holes in asymptotically flat spacetimes, Ashoke Sen and collaborators have carried out a systematic program of reproducing the logarithmic corrections as one-loop effects in quantum supergravity \cite{Banerjee:2010qc,Banerjee:2011jp,Sen:2011ba}. Extending such an impressive program to the class of asymptotically AdS black holes would clarify many issues of quantum gravity in asymptotically AdS spacetimes. A first attempt at reproducing the coefficient of the $\log N$ term  for the magnetically charged asymptotically $AdS_4\times S^7$ black holes whose microscopic entropy was discussed in \cite{Benini:2015eyy} was presented simultaneously in   \cite{Liu:2017vll} and \cite{Jeon:2017aif}; ultimately the correct computation was presented in \cite{Liu:2017vbl} and relied on contributions coming from the asymptotically $AdS_4$ region rather than the near-horizon region.  Our main result in this manuscript, the presence of a term of the form $(g-1)(7/18)\log N$ in the expansion of the topologically twisted index, is a natural target to be reproduced from quantum supergravity in massive type  IIA theory.

The manuscript is organized as follows. In section \ref{Sec:Fieldtheory} we briefly review the ingredients of the topologically twisted index and its large-$N$ evaluation. Section \ref{Sec:Numerical} presents our results as obtained from a direct numerical analysis of the index, with emphasis on the parts of the numerical analysis that we are able to reproduce analytically in section \ref{Sec:Analytical}. Section  \ref{Sec:Analytical} presents various analytical insights into the sub-leading corrections to the index. In particular, we are able to precisely track the various contributions of each of the components of the index and demonstrate certain cancellations explicitly. We include some brief and preliminary comments about the holographic side in section \ref{Sec:Holography} and conclude in section \ref{Sec:Conclusions}. Some background on the Euler-Maclaurin formula for polytopes, repeatedly used in the main text, is relegated to appendix \ref{App.EM}.


\section{The topologically twisted index for the dual of Massive IIA}\label{Sec:Fieldtheory} 

In this section we briefly review the construction of the topologically twisted index for  ${\cal N}=2$ field theories on a genus $g$ Riemann surface times a circle,  $\Sigma_g \times S^1$, following the standard literature \cite{Benini:2015noa,Benini:2016hjo,Closset:2016arn}. The field theory is defined on  a manifold with background metric 
\be
ds^2 =ds^2(\Sigma_g)+\beta^2\mathrm{d}t^2, 
\ee
where $t$ parametrizes a circle of radius $\beta$. We also consider the background field  $A^R$ which effectively implements the topological twist.  The  gauge sector to be localized contains  Yang-Mills and  Chern-Simons components described by the Lagrangians ${\cal L}_{YM}$ and ${\cal L}_{CS}$. One can also consider a set of flavor symmetries characterized by their Cartan-valued magnetic background
\be
J^f=\frac{1}{2\pi}\int\limits_{\Sigma_g}F^f = \vec{n}.
\ee
There are fugacities naturally associated with flavor and dynamical fields as:
\be
\label{Eq:fugacities}
y=\exp\left[ i \left(A_t^f+i\beta \sigma^f\right)\right], \qquad x=\exp\left[ i \left(A_t+i\beta \sigma\right)\right],
\ee
where the constant potential $A_t^f$ is a flat connection for the flavor symmetry and $\sigma^f$ is a real mass for the three-dimensional field theory. The topologically twisted index is generically of the form:
\be
Z(\vec{n}, y)=\frac{1}{|W|}\sum\limits_{m\in \Gamma_{\mathfrak h}}\; \oint\limits_{\cal C}Z_{int}(x,y;m,\vec{n}),
\ee
where $|W|$ is the dimension of the Weyl group. In the expression above one sums over all the magnetic fluxes in the co-root lattice $\Gamma_{\mathfrak h}$ of the gauge group and integrates over the Jeffrey-Kirwan  contour ${\cal C}$. 

Let us now recall  the building blocks that go into $Z_{int}$ in more detail. The contribution of a chiral multiplet is 
\be
Z_{\mathrm{1\text-loop}}^{\mathrm{chiral}}=\prod\limits_{\rho \in {\cal R}}\left(\frac{x^{\rho/2}y^{\rho_f/2}}{1-x^\rho y^{\rho_f}}\right)^{\rho(m)+\rho_f(n)-q+1},
\ee
where ${\cal R}$ is the representation of the gauge group $G$, $\rho$ denote the weights, $q$ is the R-charge of the field and $\rho_f$ is the weight of the multiplet under the flavor symmetry.  The contribution of the gauge multiplet through the one-loop determinant is:
\be
Z_{\mathrm{1\text-loop}}^{\mathrm{vector}}=\prod\limits_{\alpha \in G}(1-x^\alpha)\left(id u\right)^r,
\ee
where $r$ is the rank of the gauge group, $\alpha$ denotes roots of $G$.  We also used  $u=A_t+i\beta \sigma $ which lives on the complexified Cartan subalgebra  and is simply related to $x$ in (\ref{Eq:fugacities}) by $x=e^{iu}$. The only classical contribution in the theories we consider comes from the Chern-Simons term 
\be
Z_{\mathrm{classical}}^{CS}=x^{km}, 
\ee
where $k$ is the Chern-Simons level and $m$ is a magnetic flux originating as a solution of the localization locus.  Combining the above factors then gives the explicit form of the topologically twisted index
\be
Z(\frak{n}, y) =\frac{1}{|W|}\sum\limits_{m\in \Gamma_{\mathfrak h}}\oint\limits_{\cal C}\prod\limits_{{\rm Cartan}}\left(\frac{dx}{2\pi i \, x}x^{k\frak{m}}\right)
\prod\limits_{\alpha\in G}(1-x^\alpha)\prod\limits_{I}\prod\limits_{\rho_I\in \mathfrak{R}_I}\left(\frac{x^{\rho_I/2}y_I}{1-x^{\rho_I} y_I}\right)^{\rho_I(\frak{m})-\frak{n}_I+1},
\ee
where $\alpha$ are the roots of $G$ and $\rho_I$ are the weights of the representation $\mathfrak{R}_I$ and $m$ are gauge magnetic fluxes living in the co-root lattice $\Gamma_{\mathfrak h}$, and the integration is over the Jeffrey-Kirwan contour. For more explicit examples see \cite{Benini:2015noa,Benini:2016hjo,Closset:2016arn,Hosseini:2016tor,Hosseini:2016ume}.

\subsection{The leading $N^{5/3}$ behavior of the topologically twisted index}

The field theory of interest, first identified in \cite{Guarino:2015jca}, is an ${\cal N}=2$ supersymmetric Chern-Simons gauge theory with gauge group $SU(N)$ at level $k$ which is coupled to three chiral multiplets in the adjoint representation denoted by $X,Y$ and $Z$. The theory contains a superpotential $W={\rm Tr}\, X[Y,Z]$. The global symmetry of this theory is $SU(3)\times U(1)_R$, and the maximal torus of this symmetry is described by three $U(1)$'s, two of which correspond to flavor symmetries.  Earlier discussions of supersymmetric Chern-Simons matter theory were presented in \cite{Nishino:1991sr,Gates:1991qn}. 

The three-dimensional field theory is placed on the product of a genus $g$ Riemann surface and a circle, $\Sigma_g\times S^1$, with a topological twist along the Riemann surface $\Sigma_g$.  We further introduce background magnetic charges $\mathfrak n_a$, $a=1,2,3$ satisfying the constraint $\sum_{a=1}^3\mathfrak n_a=2(g-1)$ demanded by supersymmetry. The final expression for the topologically twisted index is then \cite{Benini:2017oxt,Azzurli:2017kxo,Hosseini:2017fjo}:
\begin{equation}
\label{Eq:Index}
Z=\frac{1}{N!}\prod\limits_{a=1}^{3}\frac{y_a^{N^2(1+\mathfrak n_a-g)/2}}{(1-y_a)^{N(1+\mathfrak n_a-g)}}\sum\limits_{I\in BAE}
\left(\det\mathbb{B}\right)^{g-1}\prod\limits_{i\neq j}\left[\left(1-\frac{x_i}{x_j}\right)^{1-g}\prod\limits_{a=1}^3\left(1-y_a\frac{x_i}{x_j}\right)^{g-1-\mathfrak n_a}\right].
\end{equation}
Here the fugacities are constrained by $\prod_{a=1}^3y_a=1$.
We will numerically explore the index in the genus zero case, but most of the results generalize immediately to the arbitrary genus case.  The Bethe Ansatz Equations (BAE) corresponding to the position of the poles following the Jeffrey-Kirwan prescription are: 
\be
\label{Eq:BAEprecursor}
e^{iB_i(x)}=x_i^k\prod\limits_{j\neq i}^N\prod\limits_{a=1}^3\frac{x_i -y_ax_j}{x_j-y_ax_i}=1.
\ee
The determinant of the matrix ${\mathbb{B}}$ is the Jacobian of the transformation  $(x_i\to B_i(x))$:
\be
\mathbb{B}_{ij}=\frac{\partial e^{iB_i(x)}}{\partial \log x_j}.
\ee
A more explicit and convenient form for the matrix  ${\mathbb{B}}$   is
\be
\label{Eq:MatrixB}
\mathbb{B}_{ij}=e^{iB_i(x)}\bigg[\left(k+\sum\limits_{l=1}^ND_{il}\right)\delta_{ij}-D_{ij}\bigg], 
\ee
where
\be
D_{ij}=z\frac{\partial}{\partial z}\log \left(\frac{z-y_1}{1-y_1z}\; \frac{z-y_2}{1-y_2z}\; \frac{z-y_3}{1-y_3z}\right).
\ee

Recalling the definitions: $x_i=e^{iu_i}$ and $y_a=e^{i\Delta_a}$, the BAE's in (\ref{Eq:BAEprecursor}) can be rewritten as 
\be
\label{Eq:BAE-u}
ku_i +i\sum\limits_{j=}^N\sum\limits_{a=1}^3\bigg[\Li_1(e^{i(u_j-u_i+\Delta_a)})-\Li_1(e^{i(u_j-u_i-\Delta_a)})\bigg]-2\pi n_i +\pi N=0,
\ee
where $\Li_1(z)=-\log(1-z)$, and the set of integers $\{n_i\}$ characterize the ambiguity implicit in taking the log of (\ref{Eq:BAEprecursor}).  The above equations can be obtained from the so-called Bethe potential:
\be
{\cal V}(u_i)=-\sum\limits_{i=1}^N\frac{k}{2}u_i^2 +\frac{1}{2}\sum\limits_{i,j=1}^N\sum\limits_{a=1}^3
\bigg[{\rm Li}_2(e^{i(u_j-u_i+\Delta_a)})-{\rm Li}_2(e^{i(u_j-u_i-\Delta_a)})\bigg]+\sum\limits_{i=1}^N 2\pi n_i u_i.
\label{eq:Bethe}
\ee
For simplicity, we limit ourselves to the domain 
\be
0< \Delta_a < 2\pi, \qquad \sum\limits_{a=1}^3\Delta_a=2\pi,
\ee
and take all expressions to be on their principal branches.

In the large-$N$ limit it is natural to consider a continuous distribution of eigenvalues $u_i$ denoted by  $u(t)$ where $t$ is the continuous parameter generalizing the subindex $i=1,\ldots, N$ of each solution
\begin{equation}
i\ \to\  t_i\ \to\  t(i)\ \to\  t.
\end{equation}
Along with this, we introduce the continuum eigenvalue density
\begin{equation}
\fft{di}{dt}=(N-1)\rho(t).
\label{eq:didt}
\end{equation}
The factor $N-1$ is important when working beyond leading order, so that we can normalize $\rho(t)$ according to
\begin{equation}
N-1=\int_1^Ndi=(N-1)\int_{t_1}^{t_N}\rho(t)dt\qquad\Rightarrow\qquad\int_{t_1}^{t_N}\rho(t)dt=1.
\end{equation}
Here $t_1=t_-$ and $t_N=t_+$ correspond to the left and right endpoints of the distribution, and the shift by $-1$ can also be understood as an endpoint correction, with $-1/2$ coming from each endpoint.

It was established, already in \cite{Hosseini:2016tor}  (see also \cite{Jafferis:2011zi,Fluder:2015eoa}), that the eigenvalues follow a distribution of the form
\begin{equation}
u(t)= N^{1/3}(it +v(t)).
\label{eq:urep}
\end{equation}
In terms of the functions $\rho(t)$, $v(t)$ and the Lagrange multiplier $\mu$, the leading term of the index can be written as 
\begin{align}
\frac{1}{N^{5/3}}{\cal V}(\rho, v, \mu)&=\int \mathrm{d}t \rho(t) \bigg[-ik\, t\, v(t)-\frac{k}{2}(v(t)^2-t^2)^2\bigg]
+iG_\Delta\int \mathrm{d}t\frac{\rho(t)^2}{1-i{v}'(t)}\nn\\
&\qquad-i\mu\left(\int \mathrm{d}t \rho(t)-1\right),
\label{eq:cBethe}
\end{align}
where
\be
\label{Eq:g-def}
G_\Delta=\sum\limits_{a=1}^3g_+(\Delta_a), \qquad g_+(x)=\frac{1}{6}x^3-\frac{\pi}{2}x^2+\frac{\pi^2}{3}x.
\ee
Extremization leads to the following solution for the saddle point:
\begin{align}
v(t)&= -\frac{1}{\sqrt{3}}\, t, &\mu& =\fft{9G_\Delta}{8t_*} \left(1-\fft{i}{\sqrt{3}}\right), \nonumber \\
\rho(t)&=\fft3{4t_*}\left(1-\left(\fft{t}{t_*}\right)^2\right), & 
t_{\pm}&= \pm t_*,
\label{eq:vrho}
\end{align}
where
\begin{equation}
t_*=\frac{3^{5/6}G_\Delta^{1/3}}{2k^{1/3}}.
\end{equation}
The support of the distribution is in the segment  $[t_-, t_+]$. At leading order, the eigenvalue density $\rho(t)$ vanishes at the ends of the segment.  However, at subleading order, it no longer vanishes, with $\rho(t_\pm)$ scaling as $\mathcal O(N^{-1/3})$, as we will show below.

The leading order in $N$ evaluation of the index is \cite{Benini:2017oxt,Azzurli:2017kxo,Hosseini:2017fjo}\footnote{There are some slight differences in the expressions presented by these three papers, we follow \cite{Benini:2017oxt}.}:
\be
\log Z(\mathfrak n_a, \Delta_a)=\frac{3^{7/6}}{2^{5/3}5}\left(1-\fft{i}{\sqrt{3}}\right)k^{1/3}N^{5/3}\left(\Delta_1\Delta_2\Delta_3\right)^{2/3} \sum\limits_{a=1}^3\frac{\mathfrak n_a}{\Delta_a}.
\label{eq:lZlo}
\ee
The above result has been  successful matched to the black hole entropy in massive type IIA supergravity  \cite{Benini:2017oxt,Azzurli:2017kxo,Hosseini:2017fjo}. Some original discussion about holography between massive type IIA and the dual Chern-Simons matter  field theory was presented first in \cite{Aharony:2010af}.  In subsequent works   \cite{Jafferis:2011zi,Fluder:2015eoa}  the matching of the free energy on $S^3$ for the field theory and the $AdS_4$ vacuum solution on the gravity side was presented.  Matching the Bekenstein-Hawking entropy is thus far the most impressive test of this duality pair. We hope that the work reported here will set the stage for a precision test by eventually matching sub-leading in $N$ contributions to the entropy on the field theoretic (microscopic) side to the gravitational (macroscopic) side.

\section{Numerical evidence for $-\frac{7}{18}\log N$ behavior in the index}\label{Sec:Numerical}

To explore the behavior of the topologically twisted index beyond the leading order, we begin with a numerical approach.  The general method is to fix a Chern-Simons level $k$ and pick a set of chemical potentials $\Delta_a$, and then numerically solve the BAE, (\ref{Eq:BAE-u}), for different values of $N$.  In practice, we always take $k=1$ and work with $N$ from 100 to 600 in steps of 20, with smaller values used in the initial stages for testing purposes.  Solution of the BAE yields a set of complex eigenvalues, $\{u_i\}$, which can then be inserted into (\ref{Eq:Index}) to evaluate the index.

The numerical work is performed in Mathematica, using FindRoot.  The initial vector for root finding is constructed from the leading order distribution, (\ref{eq:vrho}), and the integers $n_i$ are correspondently selected.  The call to FindRoot is iterated twice, first with WorkingPrecision set to MachinePrecision, and subsequently with WorkingPrecision set to 200.  The RMS error is monitored to check for convergence to a true solution.  To gather additional numerical insight, we split the index (\ref{Eq:Index}) into its components
\begin{equation}
Z=\fft1{N!}\sum_{I\in B AE}Z_{\rm{det}}Z_{\rm{chiral}}Z_{\rm{vector}},
\end{equation}
where
\begin{align}
Z_{\rm{det}}&=(\det\mathbb B)^{g-1},\nn\\
Z_{\rm{chiral}}&=\prod_{a=1}^3\fft{y_a^{N^2(1+n_a-g)/2}}{(1-y_a)^{N(1+n_a-g)}}\prod_{i\ne j}\left(1-y_a\fft{x_i}{x_j}\right)^{g-1-n_a},\nn\\
Z_{\rm{vector}}&=\prod_{i\ne j}\left(1-\fft{x_i}{x_j}\right)^{1-g}.
\label{eq:Zparts}
\end{align}
These individual terms correspond to the contributions from the Jacobian factor, the chiral multiplets and the vector multiplet, respectively.  Either the full index or its components can then be evaluated using the numerical solution for the eigenvalues.  Note that the $N!$ term is the dimension of the Weyl group and is canceled by summing over the permutations of the eigenvalues $\{u_i\}$.

The numerical results are presented for genus zero; while it is straightforward to extend the results to higher genus, little additional information is obtained in doing so.  Similarly, given the explicit form in which the magnetic fluxes $n_a$ enter in the index we consider results for vanishing fluxes. We start with the full index, which can be expanded in powers of $N$:
\begin{equation}
\label{Eq:fit-tot}
\Re\log Z=f_0N^{5/3}+h_1N^{2/3}+h_3N^{1/3}+h_4\log N+h_5+\cdots.
\end{equation}
Here $f_0$ is the leading order contribution given by (\ref{eq:lZlo}) with $k=1$.  The remaining coefficients, $h_i$, are functions of $\Delta_a$ and the magnetic charges $\mathfrak n_a$, with the dependence on the latter being at most linear, as evident from (\ref{eq:Zparts}).  The actual fit is a linear least squares fit to the numerical data with the known leading order term $f_0N^{5/3}$ subtracted out.  In addition to the entries in (\ref{Eq:fit-tot}), we allow for inverse powers of $N$ of the form $N^{-\ell/3}$ for $\ell=1,\ldots,6$, as they are potentially important, especially near the lower limit of $N=100$.  The result for several values of $\{\Delta_a\}$ are collected in Table~\ref{tbl:tot}.  Our key observation is that the coefficient $h_4$ of $\log N$ is approximately independent of the chemical potentials and agrees at the $10^{-4}$ level with
\begin{equation}
h_4=-\fft7{18}=-0.38888\ldots.
\end{equation}
The attentive reader would have noticed that we have omitted several terms, such as $N^{4/3}$ and $N$ as well as $N^{\ell/3}\log N$ ($\ell\ne0$) from the fit.  The asymptotic growth of the index is consistent with the first subdominant correction scaling as $N^{2/3}$, while introducing a possible $N^{1/3}\log N$ term only has a small effect on the fit, with the best fit coefficient being generally of order $10^{-5}$.  Finally, terms of the form $N^{\ell/3}\log N$ with $\ell<0$ are generally small in the large-$N$ limit, and have little effect on the dominant terms of the fit, which are the only ones we are interested in.  Moreover, we are not making any claims on the analytic structure of the expansion beyond the explicit terms presented in (\ref{Eq:fit-tot}).

\begin{table}[t]
\begin{center}
\begin{tabular}{c|rrrrr}
$\Delta_a/2\pi$&$h_1$&&$h_3$&$h_4$&$h_5$\\
\hline
$1/3,1/3,1/3$&$-0.04877$&$\hphantom{-0.00000}$&$\hphantom{-}2.42234$&$-0.38895$&$-2.15010$\\
$.26,.34,.40$&$-0.04913$&&$2.53991$&$-0.38896$&$-2.21735$\\
&$+0.00238n_1\kern-1em$&&$+0.07645n_1\kern-1em$&&$-0.04049n_1\kern-1em$\\
&$-0.00008n_2\kern-1em$&&$+0.02505n_2\kern-1em$&&$-0.01494n_2\kern-1em$\\$.2,.3,.5$&$-0.05607$&&$2.79564$&$-0.38888$&$-2.37246$\\
&$+0.00449n_1\kern-1em$&&$+0.16129n_1\kern-1em$&&$-0.08789n_1\kern-1em$\\
&$-0.00228n_2\kern-1em$&&$+0.07168n_2\kern-1em$&&$-0.04710n_2\kern-1em$\\\end{tabular}
\end{center}
\caption{The coefficients $\{h_i\}$ in the large-$N$ expansion of the index as given in (\ref{Eq:fit-tot}).  The coefficient $h_4$ of $\log N$ appears to be universal and agrees well with the value $-7/18=-0.38888\ldots$.}
\label{tbl:tot}
\end{table}

\subsection{Contribution from the determinant factor}\label{subsec:individual}

For a closer look at the numerical data, we highlight the contribution from the determinant factor in (\ref{eq:Zparts}).  Here the relevant expansion takes the form
\begin{equation}
\Re\log Z_{\mathrm{det}}=-\ft23N\log N-cN+g_2N^{1/3}\log N+g_3N^{1/3}+g_4\log N+g_5+\cdots,
\label{eq:Zdetc}
\end{equation}
where
\begin{equation}
c=\fft13\left(\log\fft{72\pi^3k}{G_\Delta}-5+i\pi\right),
\label{Eq:c}
\end{equation}
is a constant pertaining to the log in the leading term.  The $-2/3$ factor along with $c$ were initially obtained numerically, but can be confirmed by an analytic calculation, as we demonstrate in the next section.  What this highlights is that the determinant factor is unimportant as far as the leading order $N^{5/3}$ behavior of the full index is concerned, but contributes at $\mathcal O(N\log N)$.  Since this dependence is not seen in the full index, it is necessarily canceled by an opposite $\fft23N\log N+cN$ contribution from the vector and chiral multiplets, combined.  The $g_i$ coefficients are presented in Table~\ref{tbl:gi} for genus $g=0$.  By definition, they are independent of the magnetic charges $\mathfrak n_a$, but do depend non-trivially on the chemical potentials.

In addition to the leading $-\fft23N\log N$ behavior of $\Re\log Z_{\mathrm{det}}$, the numerical fit suggests the presence of a term of $\mathcal O(N^{1/3}\log N)$ which is likewise not present in the full index.  As we will see, this term arises as an $\mathcal O(N^{-2/3})$ correction to the leading term.  It then cancels in the full index for the same reason that the $\mathcal O(N\log N)$ term cancels between the vector/chiral multiplet contribution and the determinant contribution.

\begin{table}[t]
\begin{center}
\begin{tabular}{c|rrrrr}
$\Delta_a/2\pi$&&$g_2$&$g_3$&$g_4$&$g_5$\\
\hline
$1/3,1/3,1/3$&$\hphantom{-0.00000}$&$-0.04759$&$\hphantom{-}2.47102$&$-0.32681$&$-2.31278$\\
$.26,.34,.40$&&$-0.04835$&$2.52569$&$-0.32666$&$-2.34877$\\
$.2,.3,.5$&&$-0.05095$&$2.71126$&$-0.32707$&$-2.46586$\\
\end{tabular}
\end{center}
\caption{The coefficients $\{g_i\}$ for the determinant contribution to the index, (\ref{eq:Zdetc}).}
\label{tbl:gi}
\end{table}

We can gain further insight on $\Re\log Z_{\mathrm{det}}$ by noting that the $\mathbb B$ matrix in  (\ref{Eq:MatrixB}) can be broken up into a sum of two terms, $\mathbb{B}=B_d-B_e$, where
\begin{align}
	(B_d)_{ij}&=\delta_{ij}d_i,\qquad d_i=k+\sum_lD_{il},\nn\\
	(B_e)_{ij}&=D_{ij},
\end{align}
with
\begin{equation}
D_{ij}=\sum_a\left(\fft1{1-y_a x_{ij}}-\fft1{1-x_{ij}/y_a}\right).
\end{equation}
Note that $D$ is symmetric, and hence so is $\mathbb B$.  Given this split, the determinant contribution decomposes into a sum of diagonal and off-diagonal terms
\begin{equation}
	\log Z_{\mathrm{det}}=\log Z_{\mathrm{diag}}+\log Z_{\mathrm{off\text-diag}}=(g-1)\Tr\log B_d+(g-1)\Tr\log(1-B_d^{-1}B_e).
\label{eq:twoterms}
\end{equation}
Numerically, we find the $\mathbb B$ matrix to be dominated by its diagonal, which translates to $\log Z_{\mathrm{diag}}$ scaling as $N\log N$, accounting for the leading two terms in (\ref{eq:Zdetc}), and $\log Z_{\mathrm{off\text-diag}}$ scaling as $N^{1/3}$.  More precisely, we observe the behavior
\begin{align}
\Re\log Z_{\mathrm{diag}}&=-\ft23N\log N-cN+\alpha_2N^{1/3}\log N+\alpha_3N^{1/3}+\alpha_4\log N+\alpha_5+\cdots,\nn\\
\Re\log Z_{\mathrm{off\text-diag}}&=\kern15.65em\beta_3N^{1/3}+\beta_4\log N+\beta_5+\cdots,
\label{eq:albe}
\end{align}
with the coefficients $\{\alpha_i\}$ and $\{\beta_i\}$ tabulated in Table~\ref{tbl:albe}.  Note that the $\beta_4$ coefficient is consistent with $-1/3$, independent of the chemical potentials, suggesting that the off-diagonal determinant term contributes $-(1/3)\log N$ to the index.

\begin{table}[t]
\begin{center}
\begin{tabular}{c|rrrrr}
$\Delta_a/2\pi$&&$\alpha_2$&$\alpha_3$&$\alpha_4$&$\alpha_5$\\
\hline
$1/3,1/3,1/3$&$\hphantom{-0.00000}$&$-0.04755$&$-0.29796$&$\hphantom{-}0.00783$&$-0.10364$\\
$.26,.34,.40$&&$-0.04831$&$-0.29878$&$\hphantom{-}0.00800$&$-0.10645$\\
$.2,.3,.5$&&$-0.05107$&$-0.29764$&$\hphantom{-}0.00288$&$-0.08441$\\
\noalign{\bigskip}
$\Delta_a/2\pi$&&&$\beta_3$&$\beta_4$&$\beta_5$\\
\hline
$1/3,1/3,1/3$&$\hphantom{-0.00000}$&$\hphantom{-0.00000}$&$\hphantom{-}2.76839$&$-0.33352$&$-2.21523$\\
$.26,.34,.40$&&&$2.82387$&$-0.33353$&$-2.24853$\\
$.2,.3,.5$&&&$3.01069$&$-0.33334$&$-2.36314$\\
\end{tabular}
\end{center}
\caption{The coefficients $\{\alpha_i\}$ and $\{\beta\_i\}$ for the diagonal and off-diagonal determinant contributions, (\ref{eq:albe}).}
\label{tbl:albe}
\end{table}

Focusing only on the $\log N$ term in the full index, we see that it receives contributions for the vector/chiral multiplets (actually only from the vector multiplet, as we will show below) and both the diagonal and off-diagonal parts of the determinant factor.  This numerical evidence, as collected in Table~\ref{tbl:logN}, points to the diagonal part of the determinant contribution to be dependent of the values of $\{\Delta_a\}$. However, the sum of the diagonal determinant part and the vector/chiral multiplet contribution is reliably $\{\Delta_a\}$-independent and equal to $-1/18$.  In the next section, we analytically confirm the $-(1/3)\log N$ contribution coming from the off-diagonal part of the determinant.  However, obtaining analytic support for the remaining $-(1/18)\log N$ contribution remains an open issue.

\begin{table}[t]
\begin{center}
\begin{tabular}{c|rrr|r}
$\Delta_a/2\pi$&vector/chiral&det diag&det off-diag&vector/chiral $+$ diag\\
\hline
$1/3,1/3,1/3$&$-0.06627$&$0.00783$&$-0.33352$&$-0.05844$\\
$.26,.34,.40$&$-0.06642$&$0.00800$&$-0.33353$&$-0.05842$\\
$.2,.3,.5$&$-0.06132$&$0.00288$&$-0.33334$&$-0.05844$\\
\hline
&&&$-1/3$&$-1/18$
\end{tabular}
\end{center}
\caption{Contributions to $\log N$ from the vector/chiral multiplets, the diagonal terms in the determinant and the off-diagonal terms in the determinant.}
\label{tbl:logN}
\end{table}

\section{The index beyond the large-$N$ limit: Analytic results}\label{Sec:Analytical}

Having provided numerical evidence for the coefficient of the $\log N$ term in the topologically twisted index, we now turn to a partial analytic treatment of its sub-leading structure.  While a full treatment is still lacking, we provide explicit expressions for a number of sub-leading coefficients and highlight possible methods for a more systematic expansion beyond leading order.  Although this section can be read independently from the numerical work of section \ref{Sec:Numerical}, the synergy between the two is obvious.

At leading order, the topologically twisted index can perhaps be evaluated most straightforwardly from the Bethe potential, (\ref{eq:Bethe}).  In particular, after taking the large-$N$ limit, one derives the eigenvalue distribution (\ref{eq:vrho}) specified by $\rho(t)$ and $v(t)$ and then inserts this back into the continuum Bethe potential $\mathcal V(\rho,v,\mu)$ given in (\ref{eq:cBethe}).  However, in order to extend this analysis beyond the leading order, we return to the exact expression (\ref{Eq:Index}), which can be written as a product of three terms

The outline of the analytic approach is to first obtain the eigenvalues $x_i=e^{2\pi iu_i}$ beyond the leading order, and then to substitute them into the three components of the index given above.  Since we are working in the large-$N$ limit, it is still valid to use a continuum representation for the eigenvalues.  However, care must be taken when replacing sums by integrals, as a proper accounting of sub-leading terms requires the use of Euler-Maclaurin summation.  In any case, we proceed by working out the eigenvalue distribution $\rho(t)$ and $v(t)$ beyond leading order.  After that, we evaluate $Z_{\rm{det}}$, $Z_{\rm{vector}}$ and $Z_{\rm{chiral}}$, and finally put the result together.  Again, we emphasize that while we obtain the overall structure of the large-$N$ expansion, we are unable to fully evaluate all of the relevant terms below.

\subsection{Subleading corrections to the eigenvalue distribution}

We first discuss the sub-leading corrections to the eigenvalue distribution parametrized by $\rho(t)$ and $v(t)$. Our main results is an explicit derivation of the form of the corrections and the conclusion that including those sub-leading in $N$ corrections does not affect the coefficient of the $\log N$ term.  At leading order, these expressions are given by (\ref{eq:vrho}), while at sub-leading order, we must include both generic and endpoint corrections.

The natural starting point is the BAE, (\ref{Eq:BAE-u}), which determines the solutions $u_i$:
\begin{equation}
\label{Eq:BAE-N}
ku_i=-i\sum_{j,a}\log\fft{1-e^{i(u_j-u_i-\Delta_a)}}{1-e^{i(u_j-u_i+\Delta_a)}}+\pi(2n_i-N),
\end{equation}
where the integers $n_i$ account for the branch cuts of the logs.  We now take the large-$N$ limit while keeping track of sub-leading terms.  For the eigenvalues themselves, we make the continuum replacement $u_i\to u(t)$ with $u(t)$ given by (\ref{eq:urep}). While this is motivated by the large-$N$ expansion, we note that at this stage the result is still exact, provided we identify a discrete set of $t_i$'s corresponding to the $u_i$'s.

In order to deal with the sum in (\ref{Eq:BAE-N}), we use the Euler-Maclaurin formula
\begin{equation}
\sum_i f(u_i)=\int_1^Nf(u_i)di+\fft12(f(u_N)+f(u_i))+\fft1{12}\left.\fft{df(u_i)}{di}\right|_1^N+\cdots,
\end{equation}
where we have defined $f'(u_i)\equiv df(u_i)/di$.  We now transform from $i$ to $t$ using (\ref{eq:didt}). Making this change of variables then gives us
\begin{equation}
\sum_{i=1}^Nf(u_i)=(N-1)\int_{t_1}^{t_N}\!\!\rho(t)f(u(t))dt+\fft12(f(u(t_N))+f(u(t_1)))+\fft1{12(N-1)}\left.\fft1{\rho(t)}\fft{df(u(t))}{dt}\right|_{t_1}^{t_N}\!+\cdots.
\end{equation}

In the continuum limit, the BAE (\ref{Eq:BAE-N}) then becomes
\begin{align}
kN^{1/3}(it+v(t))\kern-5.5em&\nn\\
&=-i(N-1)\int_{t_--t}^{t_+-t}ds\rho(t+s)\left(\sum_a\log\fft{1-e^{N^{1/3}(-s+i(v(t+s)-v(t)))-i\Delta_a}}{e^{N^{1/3}(-s+i(v(t+s)-v(t)))}-e^{-i\Delta_a}}+i\pi\sgn(s)\right)+\cdots,
\label{eq:BAEc}
\end{align}
where we assume the eigenvalues are distributed between $t_-$ and $t_+$. The omitted terms are endpoint contributions from Euler-Maclaurin summation that are exponentially suppressed for $t$ away from the endpoints.  (The situation when $t$ is near the endpoints will be considered later.)  Expanding $\rho(t+s)$ and $v(t+s)$ in a Taylor series around $s=0$ and simultaneously re-scaling $\hat s=N^{1/3}s$ then gives
\begin{align}
k(it+v(t))&=-iN^{1/3}(1-N^{-1})\int_{N^{1/3}(t_--t)}^{N^{1/3}(t_+-t)}d\hat s\Biggl[\rho(t)\left(\sum_a\log\fft{1-e^{-\lambda\hat s-i\Delta_a}}{e^{-\lambda\hat s}-e^{-i\Delta_a}}+i\pi\sgn(\hat s)\right)\nn\\
&\kern5.5em+N^{-1/3}\Biggl(\hat s\rho'(t)\left(\sum_a\log\fft{1-e^{-\lambda\hat s-i\Delta_a}}{e^{-\lambda\hat s}-e^{-i\Delta_a}}+i\pi\sgn(\hat s)\right)\nn\\
&\kern6.5em+\fft{i}2\hat s^2\rho(t)v''(t)\sum_a\left(\fft1{1-e^{-\lambda\hat s+i\Delta_a}}-\fft1{1-e^{-\lambda\hat s-i\Delta_a}}\right)\Biggr)
+\mathcal O(N^{-2/3})\Biggr],
\label{eq:largeN}
\end{align}
up to exponentially small corrections in the bulk, where
\begin{equation}
\lambda=1-iv'(t).
\end{equation}

\subsubsection{The large-$N$ expansion in the interior}

Away from the endpoints, we may extend the integral to go from $-\infty$ to $\infty$.  The first term in (\ref{eq:largeN}) then vanishes since it is an odd function.  We are then left with a right hand side of order ${\cal O}(1)$.  Note that the even/odd split in the integrand results in an expansion within the square brackets in inverse powers of $N^{2/3}$.  Combined with the factor $(1-N^{-1})$ in (\ref{eq:largeN}), this indicates that $\rho(t)$ and $v(t)$ have an expansion in inverse powers of $N$ as well as in inverse powers of $N^{2/3}$.  Working to first subleading order, we write
\begin{equation}
\rho(t)=\rho_0(t)+N^{-2/3}\rho_1(t)+\cdots,\qquad
v(t)=v_0(t)+N^{-2/3}v_1(t)+\cdots.
\end{equation}

For simplicity, we now assume that the leading order solution has $\rho_0(t)$ quadratic in $t$ and $v_0(t)$ linear in $t$ as given in (\ref{eq:vrho}).  The BAE then reduces to
\begin{align}
k(it+v_0(t))+N^{-2/3}kv_1(t)&=-i\int_{-\infty}^\infty d\hat s\Biggl[\hat s\rho_0'(t)\left(\sum_a\log\fft{1-e^{-\lambda_0\hat s-i\Delta_a}}{e^{-\lambda_0\hat s}-e^{-i\Delta_a}}+i\pi\sgn(\hat s)\right)\nn\\
&\quad+N^{-2/3}\Biggl(\hat s\rho_1'(t)\left(\sum_a\log\fft{1-e^{-\lambda_0\hat s-i\Delta_a}}{e^{-\lambda_0\hat s}-e^{-i\Delta_a}}+i\pi\sgn(\hat s)\right)\nn\\
&\qquad+\hat s^2(\rho_0'(t)v_1'(t)+\ft12\rho_0(t)v_1''(t))\sum_a\fft{\sin\Delta_a}{\cos\Delta_a-\cosh(\lambda_0\hat s)}\Biggr)+\cdots\Biggr],
\end{align}
where now $\lambda_0=1-iv_0'(t)$.  Note that working to this order requires an understanding of certain $\mathcal O(N^{-2\alpha})$ terms that we have not written down explicitly in (\ref{eq:largeN}).  Assuming $v_0(t)=-t/\sqrt3$, so that $\lambda_0=1+i/\sqrt3$, the subleading order in the above equation leads to 
\begin{align}
kv_1(t)=-i\int_{-\infty}^\infty d\hat s\Biggl[
&\hat s\rho_1'(t)\left(\sum_a\log\fft{1-e^{-\lambda_0\hat s-i\Delta_a}}{e^{-\lambda_0\hat s}-e^{-i\Delta_a}}+i\pi\sgn(\hat s)\right)\nn\\
&\qquad+\hat s^2(\rho_0'(t)v_1'(t)+\ft12\rho_0(t)v_1''(t))\sum_a\fft{\sin\Delta_a}{\cos\Delta_a-\cosh(\lambda_0\hat s)}\Biggr].
\end{align}

As a result of direct integration  we obtain
\begin{equation}
kv_1(t)=ik\lambda_0t\fft{\rho_1'(t)}{\rho_0'(t)}+\fft{3\sqrt3}2(\rho_0'(t)v_1'(t)+\ft12\rho_0(t)v_1''(t))G_\Delta,
\label{eq:firstos}
\end{equation}
where we used the fact that $(\lambda_0)^3=8i/3\sqrt3$. We now note that all factors in this expression are real except for $i\lambda_0$ in the first term on the RHS.  As a result, this term must vanish by itself, giving $\rho_1'(t)=0$ or $\rho_1(t)=c$.  The remaining expression is a differential equation for $v_1(t)$.  Inserting the leading order eigenvalue density
\begin{equation}
\rho_0(t)=\fft3{4t_*}\left(1-\left(\fft{t}{t_*}\right)^2\right),
\label{eq:rholo}
\end{equation}
from (\ref{eq:vrho}), and demanding  $v(t)$ to be an odd function, we are thus left with the solution valid to next order
\begin{align}
\label{Eq:DensitiesCorrected}
\rho(t)=\fft3{4t_*}\left(1-\left(\fft{t}{t_*}\right)^2\right)+N^{-2/3}c_1+\mathcal O(N^{-1}),\nn\\
v(t)=-\fft{t}{\sqrt3}+N^{-2/3}c_2\fft{t}{1-(t/t_*)^2}+\mathcal O(N^{-1}).
\end{align}
These expressions hold in the interior, but not near the endpoints of the distribution, and are determined up to two constants, $c_1$ and $c_2$.  Nevertheless, we see that the constant $c_1>0$ uniformly increases the eigenvalue density at $\mathcal O(N^{2/3})$.  For $\rho(t)$ to be properly normalized, this increase must be compensated by an endpoint shift of $\mathcal O(N^{-1/3})$ so that
\begin{equation}
t_\pm=\pm t_*\left(1+\mathcal O(N^{-1/3})\right),\qquad\rho(t_\pm)=\mathcal O(N^{-1/3}).
\label{eq:epst}
\end{equation}
Note that the exact coefficient of the $\mathcal O(N^{-1/3})$ shift cannot be obtained directly from the interior solution, (\ref{Eq:DensitiesCorrected}), as endpoint corrections will contribute as well.

\subsubsection{Endpoint corrections}

As hinted at above, there are important corrections to the eigenvalue distribution localized around the end points of the interval $[t_-,t_+]$.  In order to examine these corrections, we return to the continuum form of the BAE, (\ref{eq:largeN}), and focus on the right endpoint, $t\approx t_+$
\begin{align}
k(it+v(t))&=-iN^{1/3}\int_{-\infty}^{N^{1/3}(t_+-t)}d\hat s\Biggl[\rho(t)\left(\sum_a\log\fft{1-e^{-\lambda\hat s-i\Delta_a}}{e^{-\lambda\hat s}-e^{-i\Delta_a}}+i\pi\sgn(\hat s)\right)\nn\\
&\kern3em+N^{-1/3}\Biggl(\hat s\rho'(t)\left(\sum_a\log\fft{1-e^{-\lambda\hat s-i\Delta_a}}{e^{-\lambda\hat s}-e^{-i\Delta_a}}+i\pi\sgn(\hat s)\right)\nn\\
&\kern7em+\fft{i}2\hat s^2\rho(t)v''(t)\sum_a\left(\fft1{1-e^{-\lambda\hat s+i\Delta_a}}-\fft1{1-e^{-\lambda\hat s-i\Delta_a}}\right)\Biggr)
+\cdots\Biggr]\nn\\
&\qquad-\fft{i}{2N^{1/3}}\left(\sum_a\log\fft{1-e^{-N^{1/3}\lambda(t_+-t)-i\Delta_a}}{e^{-N^{1/3}\lambda(t_+-t)}-e^{-i\Delta_a}}+i\pi\right)+\cdots,
\label{eq:epc}
\end{align}
where we have included the first endpoint correction in the Euler-Maclaurin sum.  (Since the distribution is symmetric, the endpoint correction at $t_-$ will have the same form.)

While the term in the first line of (\ref{eq:epc}) vanishes in the interior since the integrand is odd, this is no longer the case near the endpoint.  We thus have
\begin{equation}
k(it+v(t))=-iN^{1/3}\rho(t)J(N^{1/3}(t_+-t))+\cdots,
\label{eq:tpexp}
\end{equation}
where
\begin{equation}
J(T)=\int_{-\infty}^{T}\left(\sum_a\log\fft{1-e^{-\lambda_0\hat s-i\Delta_a}}{e^{-\lambda_0\hat s}-e^{-i\Delta_a}}+i\pi\right)d\hat s= 
-\sum_{a=1}^3\fft{\Li_2(e^{-\lambda_0 T+i\Delta_a})-\Li_2(e^{-\lambda_0 T-i\Delta_a})}{\lambda_0},
\end{equation}
for $T>0$.  Since $J(T)$ is of $\mathcal O(1)$ at $T=0$, but vanishes exponentially as $\mathcal O(e^{-T})$ for large $T$, we see that the endpoint correction is only important in a region of width $\mathcal O(N^{-1/3})$ near $T_+$.  Furthermore, at $t=t_+$, the consistency of (\ref{eq:tpexp}) requires that $\rho(t_+)=\mathcal O(N^{-1/3})$.

Of course, more care is needed for a complete understanding of the endpoint behavior of (\ref{eq:epc}), as the formal expansion in inverse powers of $N$ may no longer be valid when $\rho(t)$ and $v(t)$ pick up corrections in a region of width $\mathcal O(N^{-1/3})$.  We find a consistent endpoint expansion is obtained assuming
\begin{equation}
\rho(t)\sim\mathcal O(N^{-1/3}),\qquad v(t)\sim\mathcal O(1)\quad\mbox{for}\quad t\to t_+,
\end{equation}
while the derivatives are of order $d/dt\sim\mathcal O(N^{1/3})$, since the functions vary within a narrow region near the endpoints.  The continuum integral in (\ref{eq:epc}) is then of $\mathcal O(1)$, and balances the left-hand side of the expression, while the Euler-Maclaurin endpoint expansion starts at $\mathcal O(N^{-1/3})$, and with subsequent terms down by factors of $N^{-1/3}$ for each endpoint derivative.

The leading order endpoint correction may then be obtained without use of the endpoint terms in the Euler-Maclaurin sum.  However, all terms in the expansion of the integral in (\ref{eq:epc}) contribute at the same order.  Hence it would be more proper to return to the unexpanded form of the integrand as given in (\ref{eq:BAEc}).  Unfortunately, we have been unable to obtain an analytic solution to this integral equation.  Nevertheless, we see that the interior solution, (\ref{Eq:DensitiesCorrected}), picks up an endpoint correction when $|t-t_\pm|=O(N^{-1/3})$, and in this region $\rho(t)=O(N^{-1/3})$.  At the actual endpoint, we have $\rho(t_\pm)=\mathcal O(N^{-1/3})$, and the endpoint shift is also of $\mathcal O(N^{-1/3})$.  This is consistent with (\ref{eq:epst}), and the lack of an analytic result for the endpoint correction is the reason why the exact coefficients in (\ref{eq:epst}) cannot be obtained from only the interior solution.

Finally, given the scaling of the endpoint correction, we see that it leads to $\mathcal O(N^{-2/3})$ corrections to the topologically twisted index.  This is because, although the correction is of $\mathcal O(N^{-1/3})$, it is only important in a corrected region of width $\mathcal O(N^{-1/3})$.  Thus the endpoint correction is just as important as the subleading correction in the interior, (\ref{Eq:DensitiesCorrected}), which also scales as $\mathcal O(N^{-2/3})$.

\subsection{Determinant contribution}


Having looked at the subleading corrections to the eigenvalue distribution, we now turn to the individual contributions (\ref{eq:Zparts}) to the topologically twisted index.  We start with the determinant factor $\log Z_{\mathrm{det}}=(g-1)\Tr\log\mathbb B$ where the components of the matrix $\mathbb B$ are given in (\ref{Eq:MatrixB}).  Since it can be seen that $\mathbb B$ is dominated by its diagonal, we make the same split $\mathbb{B}=B_d-B_e$ as we did above for the numerical work.  The determinant contribution then breaks up into a sum of two terms, (\ref{eq:twoterms}). which we consider below.

\subsubsection{Diagonal contribution}
The diagonal contribution takes the form
\begin{equation}
\log Z_{\mathrm{diag}}=(g-1)\Tr\log B_d=(g-1)\sum_i\log d_i\qquad\mbox{where}\qquad d_i=k+\sum_lD_{il}.
\end{equation}
We start with $d_i$ and, at leading order, convert the sum over $l$ to an integral.  Thus
\begin{align}
d(t)&=k+(N-1)\sum_a\int_{t_-}^{t_+}\rho(s)ds\left[\left(1-e^{i\Delta_a}e^{i(u(t)-u(s))}\right)^{-1}-\left(1-e^{-i\Delta_a}e^{i(u(t)-u(s))}\right)^{-1}\right]+\cdots\nn\\
&=k+N\sum_a\int_{-\infty}^{\infty}\rho(s+t)ds\left[\left(1-e^{i\Delta_a}e^{{\lambda_0} N^{1/3}s}\right)^{-1}-\left(1-e^{-i\Delta_a}e^{{\lambda_0} N^{1/3}s}\right)^{-1}\right]+\mathcal O(1),
\label{eq:dZcnt}
\end{align}
where the second line holds for $t$ away from the endpoints.  Here we have used the fact that the leading order eigenvalues follow a linear distribution $u(t)=i{\lambda_0} N^{1/3}t$ with ${\lambda_0}=1+i/\sqrt{3}$.  Furthermore, since the integral is dominated near $s\approx t$, we have shifted the integration variable and extended the limits, which is valid away from the endpoints up to non-perturbative terms.

We now expand $\rho(s+t)\approx\rho(t)+s\rho'(t)+\cdots$ and make use of the fact that the expression in the square brackets is an even function of $s$ to obtain
\begin{align}
d(t)= k&+N\rho(t)\sum_a\int_{-\infty}^{\infty} ds\left[\left(1-e^{i\Delta_a}e^{{\lambda_0} N^{1/3}s}\right)^{-1}-\left(1-e^{-i\Delta_a}e^{{\lambda_0} N^{1/3}s}\right)^{-1}\right]+\mathcal O(1)\nn\\
=k&-2i\fft{N^{2/3}\rho(t)}{\lambda_0}\sum_a(\Delta_a-\pi)+{\cal O}(1),
\end{align}
where the last line holds for $0<\Delta_a<2\pi$.  Taking $\sum_a\Delta_a=2\pi$, then gives
\begin{equation}
d(t)=\fft{2\pi i N^{2/3}}{\lambda_0}\rho(t)+\mathcal O(1),
\label{eq:largeNdt}
\end{equation}
which is valid for $t$ away from the endpoints.  In a region of width $\mathcal O(N^{-1/3})$ near the endpoints, the result is instead $d(t\approx t_\pm)={\cal O}(N^{1/3})$, since $\rho(t_\pm)={\cal O}(N^{-1/3})$.  However, this result cannot simply be obtained by inserting the endpoint behavior of $\rho$ into (\ref{eq:largeNdt}), as additional terms that we have neglected in (\ref{eq:dZcnt}) will then be important.

We now compute the diagonal contribution
\begin{align}
\label{Eq:Bd}
\log Z_{\mathrm{diag}}&=(g-1)\sum_i\log d_i\nn\\
&=(g-1)\left[(N-1)\int_{t_-}^{t_+}\mathrm{d}t\rho(t)\log d(t)+\frac{1}{2}\left(\log d(t_-)+\log d(t_+)\right)+\mathcal O(N^{-1}) \right]\nn \\
&=(g-1)\Biggl[N\int_{t_-}^{t_+}\mathrm{d}t\rho(t)\log\fft{2\pi iN^{2/3}\rho(t)}{\lambda_0}+\mathcal O(N^{1/3})\Biggr]\nn \\
&=(g-1)\left[\frac{2}{3}N\log N+N \int_{t_-}^{t_+}\mathrm{d}t\rho(t)\log\fft{2\pi i\rho(t)}{\lambda_0}+\mathcal O(N^{1/3})\right].
\end{align}
The integral can be performed using the leading order expression (\ref{eq:rholo}) for $\rho$ to obtain
\begin{equation}
\log Z_{\mathrm{diag}}=(g-1)\left[\fft23N\log N+\fft13N\left(\log\fft{72\pi^3k}{G_\Delta}-5+i\pi\right)+\mathcal C N^{1/3}\log N+{\cal O}(N^{1/3})\right].
\label{eq:lZdiag}
\end{equation}
Note that the $N^{1/3}\log N$ term, with coefficient $\mathcal C$, arises from the endpoints of the integral in the last line of (\ref{Eq:Bd}).  In particular, since $\rho(t\approx t_\pm)=\mathcal O(N^{-1/3})$, we find a contribution to $\log Z_{\mathrm{diag}}$ of the form $N\Delta t\rho(t_\pm)\log\rho(t_\pm)\sim N^{1/3}\log(N^{-1/3})$.  However, the precise value of $\mathcal C$ will depend on the details of the density near the endpoints.  The diagonal determinant contribution to the index can be compared with the numerical counterpart, (\ref{eq:albe}), after taking $g=0$ for the genus.  In particular, it provides an analytic justification for the $-\frac{2}{3} N\log N$ term as well as for the coefficient $c$ in (\ref{Eq:c}).

\subsubsection{Off-diagonal contribution}

We finish up the determinant by considering the off-diagonal contribution
\begin{equation}
\log Z_{\mathrm{off\text-diag}}=(g-1)\Tr\log(1-X)=-(g-1)\sum_{n=1}^\infty\fft1n\Tr X^n,
\label{eq:Fod}
\end{equation}
where we have defined $X\equiv B_d^{-1}B_e$.  As a matrix, we have
\begin{equation}
X_{ij}=\fft1{d_i}D_{ij},
\end{equation}
where
\begin{equation}
D_{ij}=D_{ji}=\sum_a\left[\fft1{1-e^{i\Delta_a+iu_{ij}}}-\fft1{1-e^{-i\Delta_a+iu_{ij}}}\right],\qquad
d_i=k+\sum_jD_{ij}.
\end{equation}
In the large-$N$ limit, we approximate $u_{ji}\equiv u_i-u_j$ with the linear relation $u_{ij}\approx i{\lambda_0} N^{1/3}t_{ij}$, so that
\begin{equation}
D_{ij}= D(t_i,t_j)=D(t_{ij})=\sum_a\left[\fft1{1-e^{i\Delta_a-{\lambda_0} N^{1/3}t_{ij}}}-\fft1{1-e^{-i\Delta_a-{\lambda_0} N^{1/3}t_{ij}}}\right].
\label{eq:Dij}
\end{equation}
Using the large-$N$ expression (\ref{eq:largeNdt}) for  $d_i= d(t_i)$ then gives
\begin{equation}
X_{ij}= X(t_i,t_j)=\fft{\lambda_0}{2\pi iN^{2/3}}\fft1{\rho(t_i)}D(t_i,t_j).
\end{equation}

We now evaluate $\Tr X^n$ in the large-$N$ limit by replacing sums with integrals
\begin{equation}
\Tr X^n\approx\int\prod_{i=1}^n\left(N\rho(t_i)\mathrm{d}t_i\right)\prod_{i=1}^nX(t_i,t_{i+1})
=\left(\fft{{\lambda_0} N^{1/3}}{2\pi i}\right)^n\int\prod_{i=1}^n\mathrm{d}t_i\prod_{i=1}^nD(t_i-t_{i+1}),
\end{equation}
where we define $t_{n+1}\equiv t_1$.  This is basically a set of convolution integrals, although there is a zero mode, which we take to be $t_n$.  The limits of the $t_n$ integral are then from $t_-$ to $t_+$, while the others can be extended to cover $-\infty$ to $\infty$ in the large-$N$ limit.  Examination of (\ref{eq:Dij}) suggests a transformation $t_i\to t_i/({\lambda_0} N^{1/3})$ for $i=1,2,\ldots,n-1$.  The resulting expression is then
\begin{equation}
\Tr X^n\approx2t_*{\lambda_0} N^{1/3}\int\prod_{i=1}^{n-1}\mathrm{d}t_i\prod_{i=1}^n\fft{\hat D(t_i-t_{i+1})}{2\pi i},
\end{equation}
where we set $t_n=0$ in the integrand to fix the zero mode.  Here
\begin{equation}
\hat D(t)=\sum_a\left[\fft1{1-e^{i\Delta_a-t}}-\fft1{1-e^{-i\Delta_a-t}}\right].
\end{equation}
The convolution can be performed by Fourier transform.  The result is
\begin{equation}
\Tr X^n\approx2t_*{\lambda_0} N^{1/3}\int\fft{dk}{2\pi}\left(\fft{\hat D(k)}{2\pi i}\right)^n,
\label{eq:TrXn}
\end{equation}
where the Fourier transform
\begin{equation}
\hat D(k)=\int\hat D(t)e^{-ikt}dk=2\pi i\fft{\sum_a\sinh[k\pi(1-\fft{\Delta_a}\pi)]}{\sinh k\pi},
\end{equation}
can be evaluated by contour integration.  Combining this with (\ref{eq:TrXn}) and inserting into (\ref{eq:Fod}) then gives
\begin{align}
\log Z_{\mathrm{off\text-diag}}&\approx-(g-1)\left[\fft{t_*{\lambda_0} N^{1/3}}\pi\int dk\sum_{n=1}^\infty\fft1m\left(\fft{\sum_a\sinh[k\pi(1-\fft{\Delta_a}\pi)]}{\sinh k\pi}\right)^n\right]\nn\\
&=(g-1)\left[\fft{t_*{\lambda_0} N^{1/3}}\pi\int dk\log\left(1-\fft{\sum_a\sinh[k\pi(1-\fft{\Delta_a}\pi)]}{\sinh k\pi}\right)\right].
\end{align}

The $\log$ in the integrand is inconvenient, but can be removed by integration by parts.  The result is
\begin{align}
\log Z_{\mathrm{off\text-diag}}\kern-2em&\nn\\
&\approx(g-1)\!\left[t_*{\lambda_0} N^{1/3}\!\!\int k dk\fft{\sum_a\Bigl((1-\fft{\Delta_a}\pi)\cosh[k\pi(1-\fft{\Delta_a}\pi)]-\coth k\pi\sinh[k\pi(1-\fft{\Delta_a}\pi)]\Bigr)}{\sinh k\pi-\sum_a\sinh[k\pi(1-\fft{\Delta_a}\pi)]}\right]\!.
\label{eq:Fodip}
\end{align}
This integral can also be performed by contour integration, although care is needed as the integrand is not suppressed on the imaginary axis.  Poles occur on the imaginary axis when the $\coth k\pi$ factor in the numerator blows up and when the denominator has zeros.  While zeros of the denominator are generally obtained from a transcendental equation, it actually simplifies when we demand $\sum_a\Delta_a=2\pi$.  In this case, the poles in the upper half plane are located at
\begin{equation}
k=\begin{cases}in,&\mbox{residue}=in;\\
2\pi in/\Delta_a,&\mbox{residue}=-2\pi in_a/\Delta_a,\end{cases}\qquad n=1,2,3,\ldots.
\end{equation}
The sum of these residues diverge.  However, it can be regulated by adding a convergence factor $e^{i\epsilon k}$.  Perhaps, more elegantly, the zeta-function regulated sum of the residues is
\begin{equation}
\sum\mbox{res}=i\zeta(-1)\left(1-\sum_a\fft{2\pi}{\Delta_a}\right)=-\fft{i}{12}\left(1-\sum_a\fft{2\pi}{\Delta_a}\right).
\end{equation}
Multiplying this by $2\pi i$ and inserting into (\ref{eq:Fodip}) then gives
\begin{equation}
\log Z_{\mathrm{off\text-diag}}=(g-1)\left[\fft{t_*{\lambda_0}}6\left(1-\sum_a\fft{2\pi}{\Delta_a}\right)N^{1/3}+\cdots\right],
\label{eq:lZod}
\end{equation}

The next term in this expansion is of $\mathcal O(\log N)$, and its coefficient can be obtained analytically as follows.  As noted above, there is a zero-mode divergence in the large-$N$ limit, which is treated separately. Thus we focus on the regularized sum
\begin{equation}
\label{Eq:OffDiag}
	\log Z'_{\mathrm{off\text-diag}}=(g-1)\sum_{i=2}^{N}\log\lambda_i=(g-1)\left[\int_1^N\log\lambda_idi+\frac{1}{2}\log\lambda_N-\frac{1}{2}\log\lambda_1+\cdots\right],
\end{equation}
where $\lambda_1<\dots<\lambda_N$ are the ordered eigenvalues of the matrix $1-X$.  The integral, along with the zero mode contribution, corresponds to that performed above, while the remaining terms arise from the Euler-Maclaurin formula.  To bound these terms, we need to examine the smallest and largest eigenvalues, $\lambda_1$ and $\lambda_N$, respectively.  To do so, we start with the matrix elements
\begin{equation}
(1-X)_{ij}=\delta_{ij}-\fft{D_{ij}}{d_i}=k\fft{\delta_{ij}}{d_i}+\fft{\delta_{ij}(\sum_lD_{il})-D_{ij}}{d_i}.
\label{eq:1-Xmat}
\end{equation}
Note that the first term vanishes in the large-$N$ limit (for fixed $k$), either as $\mathcal O(N^{-2/3})$ away from the endpoints or as $\mathcal O(N^{-1/3})$ at the vicinity of the endpoints.  This gives rise to an approximate zero mode, $\lambda_1\approx0$, with corresponding eigenvector $\vec v_1\approx(1,\ldots,1)^T$ (so that the second term in (\ref{eq:1-Xmat}) vanishes).  On the other hand, we can verify that $\lambda_N=\mathcal O(1)$, so it does not give rise to a large correction in (\ref{Eq:OffDiag}).

We can provide a refined estimate for the smallest eigenvalue $\lambda_1$.  To do so, we analytically continue $\lambda_0$ and $i\Delta_a$ to real numbers so that the matrix $1-X$ becomes real symmetric. While the vector $\vec u=(1,\ldots,1)^T$ is not an exact eigenvector of (\ref{eq:1-Xmat}), it can nevertheless be used to provide an upper bound through the variational principle
\begin{equation}
\lambda_1\le\fft{\vec u^T(1-X)\vec u}{\|\vec u\|^2}=\fft{k}N\sum_i\fft1{d_i}=f(k,\Delta_a)N^{-2/3},
\end{equation}
for some function $f(k,\Delta)$.  To get a lower bound, we make use of Weyl's inequality 
\begin{equation}
	\lambda_1(A+B)\geq\lambda_1(A)+\lambda_1(B),
\end{equation}
where $A$ and $B$ are Hermitian matrices and the eigenvalues are ordered as above.  Taking the matrices $A$ and $B$ to be the two matrices on the right-hand side of (\ref{eq:1-Xmat}) then gives
\begin{equation}
	\lambda_1\geq\frac{k}{\max\{d_i\}}+0=\fft{k}{d(0)}=g(k,\Delta)N^{-2/3},
\end{equation}
where the maximum $d_i$ arises at the midpoint, $t=0$, and $g(k,\Delta)$ may be obtained from (\ref{eq:largeNdt}).  Since both lower and upper bounds are of $\mathcal O(N^{-2/3})$, we conclude that
\begin{equation}
	\log\lambda_1=-\frac{2}{3}\log N+{\cal O}(1),
\end{equation}
and therefore, combining (\ref{Eq:OffDiag}) with (\ref{eq:lZod}), we find
\begin{equation}
	\log Z_{\mathrm{off\text-diag}}=(g-1)\left[\frac{t_*\lambda_0}{6}\left(1-\sum_a\frac{2\pi}{\Delta_a}\right)N^{1/3}+\frac{1}{3}\log N+{\cal O}(1)\right].
	\label{eq:lZodiag}
\end{equation}
As a result, we have analytically derived the $\beta_3$ and $\beta_4$ coefficients in the expansion (\ref{eq:albe}), with numerical values given in Table~\ref{tbl:albe}.

\begin{figure}[ht]
	\centering
	\includegraphics[width=0.6\columnwidth]{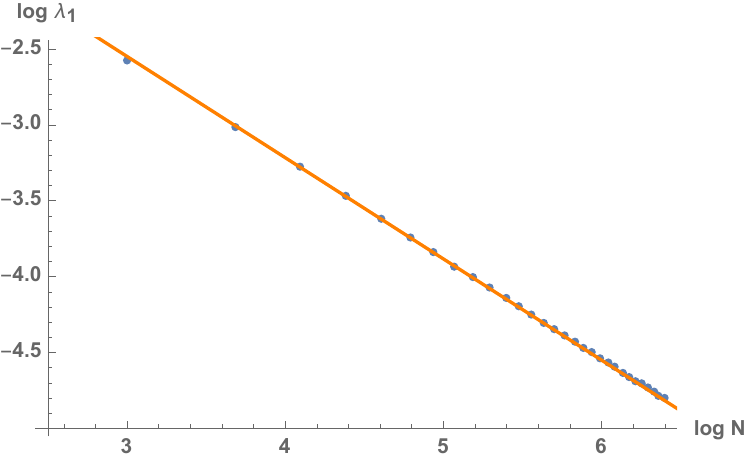}
	\caption{{\color{blue}Blue circles}: $\log\lambda_1$-$\log N$ plot for $N=20,40,\dots,600$; {\color{orange}Orange line}: $-\frac{2}{3}x+0.55$}\label{Fig:lambda1}
\end{figure}

To emphasize the synergy between our numerical and analytical approaches we provide a numerical description of the smallest eigenvalue $\lambda_1$.  Figure  \ref{Fig:lambda1} contains the  log-log plot of $\log \lambda_1$ versus $\log N$ which determines the scaling precisely.  Table \ref{Tab:lambda1} provides more details about the fit; we hope to offer the reader a flavor of the level of precision typically involved in our numerical  results.

\begin{table}
\centering
\begin{tabular}{l|rlll}

 \text{} & \text{Estimate} & \text{Standard Error} & \text{t-Statistic} & \text{P-Value}
   \\
   \hline
 $\log N$ &$ -0.66667 $& $2.43443\times10^{-8}$ &
  $ -2.7385\times 10^7 $& $2.32905\times10^{-150}$ \\
 1 &$ -0.53598$ & $2.37120\times10^{-7}$ &$ -2.26037\times
   10^6 $& $1.58663\times10^{-126} $\\
 $N^{-2/3}$ &$ -0.25391 $& 0.0000200954 & $-12635.2 $& $5.72304\times10^{-77} $\\
$ N^{-1}$ & 0.02653 & 0.000221404 &\hphantom{-}  119.822 & $1.80972\times10^{-32} $\\
$ N^{-4/3}$ &$ -0.39369$ & 0.00120808 &$ -325.884$ & $5.05952\times10^{-42}$ \\
$ N^{-5/3} $ & 0.79274 & 0.00366719 & \hphantom{-} 216.172 & $4.21395\times10^{-38}$ \\
$ N^{-2} $ &$ -1.11308 $& 0.00590413 & $-188.526$ & $8.53978\times10^{-37} $\\
$ N^{-7/3}$ & 1.12966 & 0.00393125 &\hphantom{-} 287.355 & $ 8.05402\times10^{-41} $\\ 
\end{tabular}
\caption{Fitting of $\log\lambda_1$ where $\lambda_1$ is the smallest eigenvalue as a function of $N$.} \label{Tab:lambda1}
\end{table}

\subsection{Chiral multiplet contribution}

We now turn to the chiral multiplet contribution $Z_{\mathrm{chiral}}$ given in (\ref{eq:Zparts})
\begin{equation}
\log Z_{\mathrm{chiral}}=\sum_{a=1}^3(g-1-\mathfrak n_a)\left[-\fft{N^2}2\log y_a+N\log(1-y_a)+\sum_{i\ne j}\log\left(1-y_a\fft{x_i}{x_j}\right)\right].
\end{equation}
Rewriting the last term as a restricted double sum gives
\begin{align}
\log Z_{\mathrm{chiral}}&=\sum_{a=1}^3(g-1-\mathfrak n_a)\Biggl[-N\log(1-y_a)+\sum_{\eta=\pm1}\sum_{i\le j}\log\left(1-(y_a)^\eta\fft{x_j}{x_i}\right)\nn\\
&\kern8.5em+\sum_{i\le j}\log\left(\frac{x_i}{x_j}\right)+\cdots\Biggr],
\label{eq:lZc}
\end{align}
up to a purely imaginary contribution which we do not consider.  Note that the choice of allowing $i$ to equal $j$ is convenient when translating the double sum to an integral using the Euler-Maclaurin formula.

We start with the last term in (\ref{eq:lZc})
\begin{equation}
\Lambda_0\equiv i\sum_{a=1}^3(g-1-\mathfrak n_a)\sum_{i\le j}(u_i-u_j)=i(g-1)\sum_{i\le j}(u_i-u_j).
\label{eq:L0d}
\end{equation}
At leading order in the large-$N$ expansion, we may substitute in (\ref{eq:urep}) and convert the double sum to an integral to obtain
\begin{equation}
\Lambda_0=-(g-1)\lambda_0N^{7/3}\int_{t_-}^{t_+}\rho(s)ds\int_s^{t_+}\rho(t)dt(s-t)=(g-1)\fft9{35}\lambda_0t_*N^{7/3}+\mathcal O(N^{5/3}),
\end{equation}
where the eigenvalue density $\rho(t)$ is given in (\ref{Eq:DensitiesCorrected}).  In fact, we will not need the precise form of $\Lambda_0$ as it cancels against an identical term from the vector multiplet contribution.

We now turn to the middle term in (\ref{eq:lZc}).  In the large-$N$ limit, the double sum naturally leads to a double integral, and a careful treatment can be obtained using the generalized Euler-Maclaurin formula for polytopes (see Appendix \ref{App.EM} for details).  The two-dimensional region given by $1\le i\le j\le N$ is triangular, so the Euler-Maclaurin integral decomposes as a sum of a bulk integral, boundary integral and corner contributions
\begin{align}
	\sum_{i\leq j}\log\left(1-(y_a)^\eta\frac{x_j}{x_i}\right)& = \iint\limits_{1\leq i\leq j\leq N}\log \left(1-e^{i\left(u_j-u_i+\eta\Delta_a\right)}\right)di\,dj\nonumber \\
	&\quad+ \frac{1}{2}\int_1^N \log \left(1-e^{i\left(u_N-u_i+\eta\Delta_a\right)}\right) \, di+\frac{1}{2}\int_1^N \log \left(1-e^{i\left(u_j-u_1+\eta\Delta_a\right)}\right) \, dj\nn\\
	&\quad+ \frac{1}{2}\int_1^N \log \left(1-e^{i\eta\Delta_a}\right) \, dj
	+\frac{1}{12}\int_1^N\left.\partial_j\log\left(1-e^{i(u_j-u_i+\eta\Delta_a)}\right)\right|_{j=i}di\nn\\
	&\quad+\frac{1}{4}\log \left(1-e^{i\left(u_N-u_1+\eta\Delta_a\right)}\right)+\frac{3}{4} \log \left(1-e^{i\eta\Delta_a}\right)+O(1).
\label{eq:EMlchiral}
\end{align}
The result is dominated by the bulk integral
\begin{align}
I_{\mathrm{bulk}}&\equiv\iint\limits_{1\leq i\leq j\leq N}\log \left(1-e^{i\left(u_j-u_i+\eta\Delta_a\right)}\right)di\,dj\nn\\
&=(N-1)^2\int_{t-}^{t+}\rho(s)ds\int_0^{t_+-s}\rho(s+t)dt\log\left(1-e^{-N^{1/3}t}e^{iN^{1/3}(v(s+t)-v(s))+i\eta\Delta_a}\right).
\label{eq:Ibulk}
\end{align}
Note that the $t$ integral is dominated by the lower endpoint.  In the large-$N$ limit, we let $t\to N^{-1/3}t$ and asymptotically expand around $t=0$ to obtain
\begin{align}
I_{\mathrm{bulk}}&=N^{5/3}\int_{t_-}^{t_+}ds\rho(s)^2\int_0^{\infty}dt\log\left(1-e^{-\lambda(s)t+i\eta\Delta_a}\right)+\mathcal O(N)\nn\\
&=-N^{5/3}\mathrm{Li}_2(e^{i\eta\Delta_a})\int_{t_-}^{t_+}ds\fft{\rho(s)^2}{\lambda(s)}+\mathcal O(N),
\end{align}
where $\lambda(s)=1-iv'(s)$.  Note that the $\mathcal O(N^{4/3})$ term is absent because of the symmetry properties of $\rho(t)$ and $v(t)$.  Summing over both signs of $\pm\Delta_a$ then gives
\begin{equation}
\sum_{\eta=\pm}I_{\mathrm{bulk}}=-N^{5/3}g_+'(\Delta_a)\int_{t_-}^{t_+}ds\fft{\rho(s)^2}{\lambda(s)}+\mathcal O(N).
\end{equation}

The first two boundary integrals in (\ref{eq:EMlchiral}) can be evaluated and shown to be of $\mathcal O(N^{1/3})$.  To do so, it is important to realize that the endpoint eigenvalue density scales as $\rho(t_-)=\rho(t_+)\sim\mathcal O(N^{-1/3})$.  The first integral along the diagonal gives $N\log(1-y_a)$ plus a purely imaginary term that cancels the first term in (\ref{eq:lZc}), while the second one gives a contribution of $\mathcal O(N^{1/3})$.  Finally, the first corner contribution is exponentially small, while the second one is of $\mathcal O(1)$.  Putting everything together then gives
\begin{equation}
\log Z_{\mathrm{chiral}}=\Lambda_0-N^{5/3}\int_{t_-}^{t_+}ds\fft{\rho(s)^2}{1-iv'(s)}\sum_{a=1}^3(g-1-\mathfrak n_a)g_+'(\Delta_a)+\mathcal O(N).
\end{equation}
Substituting in the leading order eigenvalue distribution then gives
\begin{equation}
\log Z_{\mathrm{chiral}}=\Lambda_0-N^{5/3}\fft3{5\lambda_0t_*}\sum_{a=1}^3(g-1-\mathfrak n_a)g_+'(\Delta_a)+\mathcal O(N).
\label{eq:lZchiral}
\end{equation}




\subsection{Vector multiplet contribution}

The final component of the topologically twisted index is the vector multiplet contribution of (\ref{eq:Zparts})
\begin{equation}
\log  Z_{\text{vector}}=-(g-1)\sum _{i\neq j} \log \left(1-\frac{x_i}{x_j}\right)=-(g-1)\left[2\sum _{i<j} \log \left(1-\frac{x_j}{x_i}\right)+\sum _{i<j} \log \left(-\frac{x_i}{x_j}\right)\right].
\end{equation}
Up to a purely imaginary term, the second sum is identical with that of (\ref{eq:L0d}) for the chiral multiplet.  Hence
\begin{equation}
\log Z_{\mathrm{vector}}=-\Lambda_0-2(g-1)\sum _{i<j} \log \left(1-\frac{x_j}{x_i}\right)+(\mbox{imaginary}).
\end{equation}
Note that here it is invalid to extend the sum to include the case $i=j$ since the expression would be logarithmically divergent.  This divergence is the origin of $\log N$ terms in the present case that are absent from the chiral multiplet contribution.

In fact, the vector multiplet sum is similar to the corresponding chiral multiplet sum in (\ref{eq:lZc}), except that we must take $y_a\to1$ and avoid the case $i=j$.  In this case, the generalized Euler-Maclaurin formula (\ref{eq:genEMl}) gives
\begin{align}\label{Eq:Vector}
	\sum_{i<j} \log \left(1-e^{i\left(u_j-u_i\right)}\right) &= \int _2^N\int _1^{j-1}\log \left(1-e^{i\left(u_j-u_i\right)}\right)di\,dj+\nonumber \\
	&\quad+\fft12 \int_1^{N-1} \log \left(1-e^{i\left(u_N-u_i\right)}\right) \, di+\fft12\int_2^N \log \left(1-e^{i\left(u_j-u_1\right)}\right) \, dj\nn\\
	&\quad+\fft12 \int_2^N\! \log \left(1-e^{i\left(u_j-u_{j-1}\right)}\right) \, dj
	+\frac{1}{12}\int_2^N\!\!\left.\partial_j\log\left(1-e^{i(u_i-u_j)}\right)\right|_{j=i-1}\!di\nn\\
	&\quad+\frac{1}{3} \log \left(1-e^{i\left(u_N-u_{N-1}\right)}\right)+\frac{5}{12} \log \left(1-e^{i\left(u_2-u_1\right)}\right)+{\cal O}(1).
\end{align}
Again, the bulk integral gives the dominant contribution, and can be rewritten as
\begin{align}
	 \int _2^N\int _1^{j-1}\log \left(1-e^{i\left(u_j-u_i\right)}\right)di\,dj \kern-10em&\nn \\
	&=\iint\limits_{1\le i\le j\le N}\log \left(1-e^{i\left(u_j-u_i\right)}\right) - \int_0^1d\ell\int_{\ell+1}^N dj\log\left(1-e^{i\left(u_j-u_{j-\ell}\right)}\right).
\label{eq:bulk12}
\end{align}
The first term is simply $I_{\mathrm{bulk}}$ for the chiral multiplet contribution, (\ref{eq:Ibulk}), evaluated at $\Delta_a=0$, while the second term gives rise to a log correction arising near the diagonal $i=j$.

For the second term in (\ref{eq:bulk12}), and more generally, terms along the diagonal, we may expand
\begin{equation}
u_j-u_{j-\ell}=\ell u'(t_j)\fft{dt_j}{dj}+\cdots,
\end{equation}
where the prime indicates a derivative with respect to $t$.  We now substitute (\ref{eq:didt}) and use (\ref{eq:urep}) to obtain, at leading order
\begin{equation}
u_j-u_{j-\ell}=\ell\fft{u'(t_j)}{N\rho(t_j)}+\cdots=i\ell\fft{\lambda(t_j)}{N^{2/3}\rho(t_j)}+\cdots.
\end{equation}
As a result, we have
\begin{align}
\int_0^1d\ell\int_{\ell+1}^N dj\log\left(1-e^{i\left(u_j-u_{j-\ell}\right)}\right)\kern-12em&\nn\\
&=\int_0^1d\ell\left[\int_1^Ndj-\int_1^{\ell+1}dj\right]\log\left(\fft{\ell\lambda(t_j)}{N^{2/3}\rho(t_j)}\right)+\mathcal O(1)\nn\\
&=(N-1)\int_{t_-}^{t_+}\rho(t)dt\left[\log\left(\fft{\lambda(t)}{N^{2/3}\rho(t)}\right)-1\right]
-\int_0^1d\ell\int_1^{\ell+1}dj\log\left(\fft{\ell\lambda(t_j)}{N^{2/3}\rho(t_j)}\right)+\mathcal O(1)\nn\\
&=-\fft23N\log N+N\int_{t_-}^{t_+}\rho(t)dt\left[\log\left(\fft{\lambda_0}{\rho(t)}\right)-1\right]+\mathcal O(N^{1/3}).
\label{eq:dint}
\end{align}
The $\mathcal O(N^{1/3})$ term comes from $\mathcal O(N^{-2/3})$ corrections in the eigenvalue distribution.  As a result, the bulk integral takes the form
\begin{align}
	\int _2^N\int _1^{j-1}\log \left(1-e^{i\left(u_j-u_i\right)}\right)di\,dj\kern-5em&\nn\\
	&=I_{\mathrm{bulk}}\bigg|_{\Delta_a=0}+
\fft23N\log N-N\int_{t_-}^{t_+}\rho(t)dt\left[\log\left(\fft{\lambda_0}{\rho(t)}\right)-1\right]+\mathcal O(N^{1/3}).
\label{eq:bildo}
\end{align}

As in the chiral multiplet case, the non-diagonal boundary integrals in (\ref{Eq:Vector}) give contributions of $\mathcal O(N^{1/3})$.  However, the diagonal boundary integrals give potentially large terms because of the $i=j$ divergence.  The first one is evaluated as in (\ref{eq:dint}), but without the $\ell$ integral
\begin{equation}
\fft12\int_2^N\log\left(1-e^{i(u_j-u_{j-1})}\right)dj=-\fft13N\log N+\fft12N\int_{t_-}^{t_+}\rho(t)dt\log\left(\fft{\lambda_0}{\rho(t)}\right)+\mathcal O(N^{1/3}),
\end{equation}
while the second one gives a term of $\mathcal O(N)$.  Finally, combining the bulk integral (\ref{eq:bildo}) with the relevant diagonal ones, and noting that the corner contributions are of $\mathcal O(\log N)$, gives
\begin{equation}
\log Z_{\mathrm{vector}}=-\Lambda_0-2(g-1)\biggl[I_{\mathrm{bulk}}\bigg|_{\Delta_a=0}+\fft13N\log N-\fft12N\int_{t_-}^{t_+}\rho(t)dt\log\left(\fft{\lambda_0}{\rho(t)}\right)+\mathcal O(N)\biggr].
\label{eq:lZvector}
\end{equation}
Although the $t$ integral in this expression can be absorbed in the term of $\mathcal O(N)$, we have chosen to keep it separate to highlight that it has the same form as that of (\ref{Eq:Bd}) in the diagonal contribution to the determinant, $\log Z_{\mathrm{diag}}$.  In particular, this demonstrates that the $\mathcal C N^{1/3}\log N$ term in (\ref{eq:lZdiag}), which originates from the same integral, exactly cancels against a similar contribution from the vector multiplet.

\subsection{The full index}

When we combine all the contributions to the index, (\ref{eq:lZdiag}), (\ref{eq:lZodiag}), (\ref{eq:lZchiral}) and (\ref{eq:lZvector}), the leading $\Lambda_0\sim\mathcal O(N^{7/3})$ term drops out, leaving
\begin{align}
\log Z&=N^{5/3}\fft3{5\lambda_0t_*}\sum_{a=1}^3(g-1-\mathfrak n_a)\left(g_+'(0)-g_+'(\Delta_a)\right)\nn\\
&\qquad+h_0N+h_1N^{2/3}+h_3N^{1/3}+h_4\log N+h_5+\cdots\nn\\
&=f_0N^{5/3}+h_0N+h_1N^{2/3}+h_3N^{1/3}+h_4\log N+h_5+\cdots,
\end{align}
where the leading order contribution, $f_0$, is the coefficient of $N^{5/3}$ in (\ref{eq:lZlo}).  As mentioned above, although $\mathcal O(N^{1/3}\log N)$ terms are present in the diagonal determinant and vector multiplet contributions, they cancel when combined.  While we have been unable to compute the precise values of the $h_i$ coefficients, we have nevertheless confirmed the analytic form of the expansion as a series in inverse powers of $N$ and $N^{1/3}$ along with $\log N$.  Note that the numerical evidence presented in Sec.~\ref{Sec:Numerical} strongly suggests the vanishing of $h_0$.  The $\mathcal O(N)$ term receives contributions from the chiral and vector multiplets as well as the diagonal part of the determinant, and it appears to cancel when all three contributions are combined.  Assuming this to be the case, we would then be left with (\ref{Eq:fit-tot}) for the large-$N$ expansion of the index.

It would of course be desirable to provide analytic expressions for the non-vanishing $h_i$ coefficients.  In order to do so, however, we would have to extend the above analysis, beginning with a more careful treatment of the endpoint corrections.  Although $h_1$ and $h_3$ appear quite challenging to pin down, there remains the possibility that the coefficient $h_4$ of $\log N$ will be amenable to a detailed study.  Since the expression in (\ref{eq:lZc}) for the chiral multiplet contribution does not contain any logarithmically divergent quantities, it will not give rise to any $\log N$ factors.  As a result, the coefficient $h_4$ will only depend on the diagonal and off-diagonal parts of the determinant and the vector multiplet contribution.  Comparison with the numerical results shown in Tbl.~\ref{tbl:logN} confirms the analytic result $h_{4\,\mathrm{off\text-diag}}=\fft13(g-1)$ from (\ref{eq:lZodiag}), while it remains to be seen whether the numerical evidence for $h_{4\,\mathrm{diag}}+h_{4\,\mathrm{vector}}=\fft1{18}(g-1)$ can be derived analytically.

\section{Comments on the holographic side}\label{Sec:Holography}

To start clarifying the  meaning of the result on the gravity side we first turn to the AdS/CFT dictionary for the case of massive type IIA theory. 
The crucial intuition for AdS/CFT in the context of massive type IIA  was proposed in \cite{Gaiotto:2009mv}, where it was argued that the Roman's mass is related to the Chern-Simons level as
\be
F_{(0)}=\frac{k}{2\pi l_s}.
\ee

A number of papers have explored the field theory  and the holographic side   while contributing to establishing the AdS/CFT dictionary in this context   \cite{Gaiotto:2009mv,Gaiotto:2009yz}. In particular,  some explicit solutions have been discussed \cite{Aharony:2010af,Petrini:2009ur,Lust:2009mb}. Other aspects of the dictionary that transpired from the above works are  as follows:
\be
\frac{R}{l_s} \sim \left(\frac{N}{k}\right)^{1/6}, \qquad g_s \sim 1/ (N^{1/6} k^{5/6}).
\ee

The supergravity regime requires, therefore that $N\gg k$.  Crucially, if we take $k=1$ we recover the same rough relation between the radius of the gravity solution and the rank of the gauge group  as in  ABJM: $R \sim N^{1/6}$. 

The complete holographic renormalization that allows the precise matching of the topologically twisted index with the black hole entropy has not been yet implemented for the massive IIA theory. It is quite reasonable to think that it follows steps similar to the analysis of magnetically charged black holes dual to topologically twisted ABJM theory performed in \cite{Halmagyi:2017hmw,Cabo-Bizet:2017xdr}. However, it would be useful to see explicitly the connection between the gravity and field theory quantities as derived in \cite{Cabo-Bizet:2017xdr}.

It would be quite interesting to provide an analytic expression for the first next-to-leading corrections of order $N^{2/3}$. Following the guidance of the ABJM/$AdS_4$ duality  \cite{Aharony:2008ug}, this term corresponds to higher curvature corrections on the dual gravity theory \cite{Bergman:2009zh,Aharony:2009fc,Drukker:2011zy}. Thus, an important result within the reach of field theory is a prediction for the contribution of higher curvature terms to the black hole entropy in  massive type IIA string theory on $AdS_4\times S^6$. It is worth pointing out that in the context of asymptotically flat black holes a complete understanding of the contributions of certain higher curvature corrections to the entropy on the gravity side has been shown to match precisely the microscopic side (see the reviews \cite{Mohaupt:2005jd,Sen:2007qy}).  The peculiarity of the situation in ABJM is even more enticing as the  ${\cal O}(N^{1/2})$ term does not follow from a standard loop expansion of supergravity, which
would be given in powers of the 11-dimensional Newton constant, $G_{11} \sim N^{ −3/2}$.  Instead, it
arises as a quantum correction in M-theory, and in particular from a shifted relation between
ABJM and M-theory parameters resulting from the eight-derivative $C_3R^4$ term where $C_3$ is the M-theory three-form gauge potential and $R^4$ schematically denotes powers of the curvature two-form (see \cite{Bergman:2009zh} for details). More precisely, one adds to the action the term $C_3\wedge I_8$, where $I_8$ is the 8-form anomaly polynomial defined in terms of Pontryagin classes  \cite{Duff:1995wd}.  We hope to attack this fascinating problem elsewhere. 

As a parting comment on the gravity side we provide a back-of-the-envelope evaluation of a  particular contribution to the logarithmic term. Recall that the main contribution in the context of quantum supergravity corrections to black holes dual to ABJM was a 2-form zero mode of $AdS_4$ considered in \cite{Liu:2017vbl}. The contribution to the partition function was shown to be  \cite{Liu:2017vbl}:
\be
\log Z_{\mathrm{1\text-loop}} =(2-\beta_2)n_2^0 \log R.
\ee
Taking into consideration that we are in a ten-dimensional theory, we have $\beta_2=3$ and using, as in \cite{Liu:2017vbl},  that $n_2^0=2(1-g)$ we find 
\be
\log Z_{\mathrm{1\text-loop}} =(2-3)2(1-g)\log R=-2(1-g)\log R = -\frac{1-g}{3} \log N.
\ee
This is but a term in the full partition function. We warn the reader that the  full computation of the partition includes all the modes and should be done carefully as we are in an even-dimensional spacetime; the computation of \cite{Liu:2017vbl}  took crucial advantage of the fact that in odd-dimensional spacetimes the Seeley-DeWitt coefficients in the heat kernel expansion vanish identically. Nevertheless, we feel encouraged to see that a known zero mode in the solution contributes a term that is compatible with the final answer. 

Curiously, in previous sections we demonstrated that the contribution to the logarithmic term coming from the off-diagonal determinant part of the index was precisely $((g-1)/3)\log N$. We have no justification for a direct identification of the contribution $ ((g-1)/3)\log N$  from the off-diagonal determinant modes in the index with the one from the zero mode on the gravity side, however tantalizing the analogy. 

Still, the prospects of a direct relationship between the elements in the topologically twisted index such as the contribution of the  vector multiplet degrees of freedom and the contribution of the gravitational degrees of freedom such as the zero mode contribution is worth exploring as it opens and extraordinary chapter in the AdS/CFT dictionary. Indeed, there is some precedent to such identification in the context of two-dimensional gravity and matrix models wherein the off-diagonal degrees of freedom in the matrix model have been shown to be crucial in reproducing aspects of the entropy at leading order \cite{Kazakov:2000pm}.

\section{Conclusions}\label{Sec:Conclusions}

In this manuscript we have provided a systematic sub-leading analysis of the topologically twisted index of supersymmetric $SU(N)$ level $k$ Chern-Simons theory coupled to matter.  There is one technical aspect in the analysis of theories with $N^{5/3}$ growth in their  degrees of freedom which seems promising. Namely, the problem of the tails in the eigenvalue distributions  and corresponding insidious mixing of orders of $N$ in the expansion of the index for theories with $N^{3/2}$ growth, as pointed out first in \cite{Benini:2015eyy} and elaborated upon in \cite{Liu:2017vll}, is ameliorated for theories with $N^{5/3}$ growth. This situation alone should help understand the full structure of corrections better and makes the corresponding field theories a particularly robust playground to explore aspects of the sub-leading in $N$ expansion of the topologically twisted index.

 It would be quite illuminating  to understand our results from a more analytic point of view. A natural starting point is  by pursuing the relation between the Bethe Potential  and the expectation value of the free energy on $S^3$ as pointed out in \cite{Hosseini:2016tor} but beyond the leading order in $N$. There are other more formal arguments establishing a relation between the topologically twisted index in $S^2\times S^1$,  or more generally on $\Sigma_g\times S^1$, and the free energy on $S^3$  pointed out in \cite{Closset:2017zgf,Closset:2018ghr}.  It is worth highlighting that in the case of theories with $N^{3/2}$ growth, that is, ABJM-like theories, there is a general result for the exact free energy on $S^3$ \cite{Fuji:2011km,Marino:2011eh} and it takes the form of an Airy function. A similar general result for theories with $N^{5/3}$ growth is yet to be found; there are only large $N$ results \cite{Jafferis:2011zi,Fluder:2015eoa,Hosseini:2016tor}. In view of its potential relationship with the index, it would be really important to obtain exact results for the free energy on $S^3$ for such theories.

There are also a number of interesting questions on the gravity side. For example,  given the successful  attempt at unifying the holographic treatment of $F_{S^3}$ and $Z_{S^2\times S^1}$ for ABJM-like theories presented in  \cite{Toldo:2017qsh} and the recent progress for superconformal theories subject to the 3d-3d correspondence \cite{Gang:2018hjd}, one might expect a similarly clarifying framework for massive type IIA theories. One would need to explicitly construct such supergravity backgrounds whose boundary is such a circle bundle and evaluate the renormalized on-shell
action.

It would be interesting to understand better the gravity side of the logarithmic corrections. In particular, it would be good perform a one-loop counting to compare with the coefficient of the $\log N$ term.  One-loop supergravity computations are substantially harder due to the number of modes contributing. Recall that in previous efforts \cite{Liu:2017vll,Jeon:2017aif,Liu:2017vbl} the odd-dimensional computation was reduced to counting of zero modes. For massive type IIA we have to face an even-dimensional spacetime and confront the local contribution to the one-loop effective action. One potential outcome of this quantum supergravity exercise would be an understanding of the universality of the one-loop result.  Namely, how theory independent is the coefficient of the logarithmic term. There is an important precedent pointing to the fact the coefficient of $\log N$ can be universal. Indeed, this is the case for free energy of various  ${\cal N}=3$  Chern-Simons matter theories evaluated on $S^3$ \cite{Fuji:2011km,Marino:2011eh}.  The universality of this quantity from the dual supergravity point of view was beautifully elucidated in \cite{Bhattacharyya:2012ye}.  We showed in \cite{Liu:2017vbl} for the case of the topologically twisted index that  on the dual quantum supergravity  side the result relies on vanishing first Betti  number for the compactifying $X^7$ manifold.

Finally, it would be quite interesting to start exploring the structure of higher curvature corrections. A summary of the intricate developments leading to a full understanding of higher curvature corrections to the black hole entropy for supersymmetric asymptotically flat black holes  in string theory is presented in  \cite{Sen:2007qy}.  For our case, the first sub-leading entropy correction to the index is of the type $N^{2/3}$. The coefficient of this term is described by  $h_1$ in Table~\ref{tbl:tot}. Although we have not been able to find an analytic expression for it, the numerical results seem to suggest that a better treatment might be possible. We hope to report on some of these open problems in the future.

\section{Acknowledgments}
We are thankful to F. Benini, C. Closset, S. M. Hosseini, A. Klemm, C. Toldo, B. Willett and  A Zaffaroni. This
work is partially supported by the US Department of Energy under Grant No. DE-SC0007859.

\appendix

\section{The Euler-Maclaurin formula}\label{App.EM}

In this appendix we review some basic aspects of Euler-Maclaurin formula. We refer to \cite{Karshon426} for details. The goal is to clarify a step that is used profusely in the main text, that of approximating a discrete sum by an integral. 

\subsection{Classical Euler-Maclaurin formula}

The classical Euler-Maclaurin formula calculates a discrete sum in terms of an integral and a series of derivatives on the boundary. The standard form of the  formula is 
\begin{align}
	\sum_{i=m}^n f(i)&=\int_m^n f(i)di+\frac{f(m)+f(n)}{2}+\frac{1}{6}\frac{f'(n)-f'(m)}{2!}-\frac{1}{30}\frac{f'''(n)-f'''(m)}{4!}\nonumber \\
	&\quad+\dots + 
	B_{2k}\frac{f^{(2k-1)}(n)-f^{(2k-1)}(m)}{(2k)!}+R_{2k}.
\end{align}
The modern form relies on the Todd operator, formally defined as 
\begin{equation}
	\Td(S)=\frac{S}{1-e^{-S}}=1+\frac{1}{2}S+\frac{1}{12}S^2-\frac{1}{720}S^4+\cdots.
\end{equation}
Note that  $\Td(-S)$ is the generating function for the Bernoulli number $b_k$
\begin{equation}
	\Td(-S)=\sum_{k=0}^\infty\frac{b_k}{k!}S^k,
\end{equation}
and the Bernoulli number vanishes for $k=2n+1, n>0$.

Let $\Td^{(2k)}$ be the Todd operator truncated at $2k$
\begin{equation}
	\Td^{(2k)}(S)=1+\frac{1}{2}S+\frac{1}{12}S^2-\frac{1}{720}S^4+\dots+\frac{b_{2k}}{(2k)!}S^{2k}.
\end{equation}
Then the classical Euler-Maclaurin formula can be re-written as
\begin{equation}
	\sum_{i=m}^nf(i)=\left.\Td(D_1)^{(2k)}\Td(D_2)^{(2k)}\int_{m-h_1}^{n+h_2}f(x)dx\right|_{h_1=h_2=0}+R_{2k+1},
\end{equation}
where $D_i=\partial_{h_i}$ and $R_{2k+1}$ is the remainder
\begin{equation}
	R_{2k+1}=\int_m^n P_{2k+1}(x)f^{(2k+1)}(x)dx,
\end{equation}
with
\begin{equation}
	P_{2k+1}(x)=(-1)^{k+1}\sum_{n=1}^\infty\frac{2\sin 2n\pi x}{(2n\pi)^{2k+1}}.
\end{equation}
The advantage of writing the Euler-Maclaurin formula using this modern language is that it naturally  suggests generalizations in higher dimensions. 

\subsection{Euler-Maclaurin formula for regular integral polytopes}

We now extend the classical Euler-Maclaurin formula to higher dimensional lattices. 
The general idea is that we are using the integral over some region to approximate a discrete sum, and all the corrections can be derived from the integral alone. In one dimension, the formula says all the boundary terms in the expansion are exactly the derivatives with respect to the upper and lower limits of the (dilated) integral $[a-h1, b+h2]$, and  $h_1$ and $h_2$ determines how we dilate the interval in the negative/positive direction. In two or higher dimensions, the interval is replaced by some polytope and $h_1, \ldots , h_n$ describes how to dilate the polytope. If the polytope has $n$ faces, $h_i$ is proportional to how the polytope is expanded in the normal direction of the $i$-th face. In particular, $h-1$,  $h_2$ and $h_3$ describes how to expand a triangle along the three edges.

For convenience, we summarize some basic concepts about lattices and polytopes. A $n$-dimensional {\bf lattice} is a discrete additive subgroup of $\mathbb{R}^n$, considered as a $\mathbb{Z}$-module. It is {\bf integral} if the inner product of lattice vectors are all integral. The {\bf dual} of a lattice consists of vectors having integral inner product with the original lattice. If the original lattice has a basis $\{a_i\}$, then the dual lattice has a {\bf dual basis} $\left<a_i^*,a_j\right>=\delta_{ij}$. 

A {\bf polytope} is a finite intersection of half-planes and is also compact. It is {\bf integral} if all vertices are in the lattice $\mathbb{Z}^n$, is {\bf simple} if exactly $n$ edges emanate from each vertex, and is {\bf regular} if additionally the edges emanating from each vertex consists of a basis of the lattice $\mathbb{Z}^n$.

Let $\Delta$ be an integral and regular polytope, described by the half-planes 
\begin{equation}
	\left<x,u_i\right>+\lambda_i\geq 0,
\end{equation}
where $u_i$ are the dual basis. (To be more precise, at each vertex $i$, the $u_j$ defining half-planes intersecting at $i$ consist of a basis of $\mathbb{Z}^n$.) Now define $\Delta(h)$ as the polytope obtained from $\Delta$ by expanding each face, i.e.
\begin{equation}
	\left<x,u_i\right>+\lambda_i+h_i\geq 0.
\end{equation}

The exact Euler-Maclaurin formula asserts that for any polynomial $p$,
\begin{equation}\label{eEM}
	\sum_{\Delta(h)\cap \mathbb{Z}^n}p=\left.\Td(D_1)\dots \Td(D_n)\int_{\Delta(h)}p~\right|_{h=0}.
\end{equation}
The above formula has a number of forms which work for more complicated polytopes and more general summands, such as smooth functions. However, (\ref{eEM}) is sufficient for our purpose because polynomial functions are dense in the space of continuous functions over an interval.


As an application, we calculate the following summation appearing in the vector multiplet contribution:
\begin{equation}
	\sum_{1\leq i<j\leq n}f(i,j),
\end{equation}
where the summation is taken over a triangle.
The three vertices of this triangle are $(1,2)$, $(1, n)$, $(n-1, n)$, and the dual basis is 
\begin{equation}
	(-1,0),\quad (0,1),\quad \left(\ft{1}{2},-\ft{1}{2}\right).
\end{equation}
Hence the dilated polytope $\Delta(h_1,h_2,h_3)$ is a triangle with vertices
\begin{equation}
	(1-h_1, 2-h_1-h_3),\quad (1-h_1, n+h_2),\quad (n-1+h_2+h_3, n+h_2).
\end{equation}

The integral over $\Delta(h)$ can be written as
\begin{equation}
	\int_{\Delta(h_1,h_2,h_3)}=\int_{2-h_1-h_3}^{h_2+n}dj\int_{1-h_1}^{j-1+h_3}di
\end{equation}
and hence the first several terms (the ones without explicit derivatives) in the Euler-Maclaurin expansion is
\begin{align}
	\sum_{1\leq i<j\leq n}f(i,j) = &\int _2^n\int _1^{j-1}f(i,j)didj+\frac{1}{2} \int_1^{n-1} f(i,n) \, di+\frac{1}{2}\int_2^n f(1,j) \, dj+\frac{1}{2} \int_2^n f(j-1,j) \, dj+\nonumber\\
	&+\frac{1}{4}f(1,n)+\frac{1}{3} f(n-1,n)+\frac{5}{12} f(1,2)+\cdots.
\label{eq:genEMl}
\end{align}
For convenience, we also list terms containing the first order derivatives
\begin{align}
	&\frac{1}{12} \int_1^{n-1} f^{(0,1)}(i,n) \, di+\frac{1}{12} \int_2^n -f^{(1,0)}(1,j) \,
   dj+\frac{1}{12} \int_2^n f^{(1,0)}(j-1,j) \, dj+\frac{1}{24}
   f^{(0,1)}(1,n)\nonumber \\
   &+\frac{1}{24} f^{(0,1)}(n-1,n)-\frac{1}{24} f^{(1,0)}(1,n)+\frac{1}{12}
   f^{(1,0)}(n-1,n)-\frac{1}{12} f^{(0,1)}(1,2)-\frac{1}{24} f^{(1,0)}(1,2),
\end{align}
where $f^{(a,b)}$ means taking $a$-th derivative with respect to the first variable and $b$-th derivative with respect to the second variable.

\bibliographystyle{JHEP}
\bibliography{BHLocalization}

\end{document}